\documentclass[12pt]{article}
\pdfoutput=1

\usepackage{putex}
\usepackage{graphicx}
\usepackage{caption}
\usepackage{amsmath}
\usepackage{amssymb}
\usepackage{array}
\usepackage{multirow}
\usepackage{mathtools}
\usepackage{comment}
\usepackage{subcaption}
\usepackage{epstopdf}
\usepackage{enumerate}
\usepackage{cite}
\usepackage{youngtab}
\usepackage{tensor}
\usepackage{slashed}
\usepackage[aligntableaux=center]{ytableau}
\usepackage[utf8]{inputenc}
\usepackage{rotating}
\usepackage{bigfoot}
\usepackage[
      colorlinks=true,
      linkcolor=blue,
      urlcolor=blue,
      filecolor=black,
      citecolor=red,
      linktocpage=true
      ]{hyperref}

\newcommand {\be} {\begin {equation}}
\newcommand {\ee} {\end {equation}}

\newcommand {\bes} {\begin {equation*}}
\newcommand {\ees} {\end {equation*}}

\newcommand{\es}[2]{%
  \begin{equation}
    \begin{aligned}
      #2
    \end{aligned}
    \phantomsection\label{#1}%
  \end{equation}%
}

\newcommand{\Z}{\mathbb{Z}}

\newcommand{\R}{\mathbb{R}}

\newcommand{\cN}{{\mathcal N}}

\newcommand{\bea}{\begin{equation}\begin{aligned}}
\newcommand{\eea}[1]{\label{#1}\end{aligned}\end{equation}}

\newcommand{\beq}{\begin{equation}}
\newcommand{\eeq}{\end{equation}}

\def\ie{\begin{equation}\begin{aligned}}
\def\fe{\end{aligned}\end{equation}}

\newcommand{\mP}{{\mathbb P}}

\numberwithin{equation}{section}

\def\<{\langle}
\def\>{\rangle}

\newcommand{\de}{\partial}
\newcommand{\h}{h^\vee}
\newcommand{\texp}{e^{\tfrac{\pi i}{3}}}
\newcommand{\discr}{\mathbf{\Delta}}
\newcommand{\rank}{\mathtt{r}}
\newcommand{\lambdaSW}{\lambda_{\mathrm{SW}}}

\begin{document}

\preprint{}

\institution{saifr}{IFT-UNESP, ICTP-SAIFR, Rua Dr Bento Teobaldo Ferraz 271, 01140-070, Sao Paulo, Brazil}
\institution{imperial}{Blackett Laboratory, Imperial College, Prince Consort Road, London, SW7 2AZ, U.K.}
\institution{stony}{Simons Center for Geometry and Physics, SUNY, Stony Brook, NY 11794, USA}

\title{Towards Bootstrapping F-theory}

\authors{Connor Behan,\worksat{\saifr}  Shai M.~Chester,\worksat{\imperial} and Pietro Ferrero\worksat{\stony}  }

\abstract{

We consider type IIB string theory with $N$ D3 branes and various configurations of sevenbranes, such that the string coupling $g_s$ is fixed to a constant finite value. These are the simplest realizations of F-theory, and are holographically dual to rank $N$ Argyres-Douglas conformal field theories (CFTs) with $SU(2)$ and $SU(3)$ flavor groups, and Minahan-Nemeschansky CFTs with $E_6, E_7$ and $E_8$ flavor groups. We use the Seiberg-Witten curves of these theories to compute the mass deformed sphere free energy $F(m)$ at large $N$ in terms of novel matrix models with non-polynomial potentials. We show how $F(m)$ can be used along with the analytic bootstrap to fix the large $N$ expansion of flavor multiplet correlators in these CFTs, which are dual to scattering of gluons on $AdS_5\times S^3$, and in the flat space limit determine the effective theory of sevenbranes in F-theory. As a first step in this program, we use the matrix models to compute the $\log N$ term in $F(m)$ and thereby fix the logarithmic threshold in the $AdS_5\times S^3$ holographic correlator, which matches the flat space prediction.
}
\date{}

\maketitle

\tableofcontents
\newpage

\section{Introduction}\label{sec:intro}

F-theory is the study of type IIB string theory in the presence of D7 branes \cite{Morrison:1996na,Morrison:1996pp,Vafa:1996xn}. Consider a stack of D7 branes whose transverse location in the 10d spacetime is denoted by a complex coordinate $z$. Since D7 branes magnetically couple to the zero form Ramond-Ramond (RR) potential $C_0$, the backreaction of D7 branes generically generates a holomorphically varying profile for the complexified string coupling $\tau_s(z)=C_0+i/g_s$, where $g_s$ is the string coupling. Since string theory in this background has finite $g_s$, it cannot be studied using worldsheet perturbation theory, which is a small $g_s$ expansion. Instead, F-theory uses the existence of the non-perturbative $SL(2,\mathbb{Z})$ duality group \cite{Hull:1994ys,Schwarz:1993mg,Schwarz:1995dk,Sen:1994yi} acting on $\tau_s$ to derive the low energy description even when string theory is strongly coupled.\footnote{The existence of an $SL(2,\mathbb{Z})$ duality group also implies that there are different types of mutually non-local {\it sevenbranes}, of which D7 branes are just a special case. We discuss this in Section \ref{sec:theories}, and from now on we adopt the terminology sevenbranes with this meaning, as it is standard in the F-theory literature.} These F-theory compactifications have been essential in building phenomenologically viable string models \cite{Donagi:2008ca,chllt19}.

While the geometric methods of F-theory have been very successful at studying the leading low energy description (see \cite{Weigand:2018rez} for a recent review), they cannot compute any of the stringy corrections to the low energy effective theory. In this sense, F-theory is analogous to M-theory, in that we know the leading low energy effective theory, but have no systematic way of computing corrections, since there is no weakly coupled worldsheet description. In this paper, we will take the first step towards computing stringy corrections to the low energy effective descriptions of the simplest F-theory models.

We consider the class of F-theory models where $\tau_s(z)$ is independent of $z$ \cite{Sen:1996vd,Dasgupta:1996ij}. These arise from $N$ D3 branes probing the following ADE type F-theory singularities:\footnote{There is another theory with constant $\tau_s$ that has an $A_0$ type singularity, but we will not consider it because there is no gluon scattering in this theory. Note that sometimes the $A_n$ theories are referred to as $\mathcal{H}_n$ to distinguish them from an infinite family of singularities which lead to $A_n$ symmetry but a varying axio-dilaton.}
\es{sing}{
A_1\,,\,\,A_2\,,\,\,D_4\,,\,\,E_6\,,\,\,E_7\,,\,\,E_8\,.
}
At large $N$, the low energy effective theory is given by supergravity on $AdS_5\times S^5$, but with a singularity whose fixed point locus is super-Yang-Mills (SYM) on $AdS_5\times S^3$ with one of the gauge groups given in \eqref{sing}. The holographic dual \cite{Fayyazuddin:1998fb,Aharony:1998xz} of these theories are given by 4d $\mathcal{N}=2$ conformal field theories (CFTs) with the flavor symmetries in \eqref{sing}. The four-point functions of flavor multiplets in the CFT is then dual to gluons scattering on $AdS_5\times S^3$, which in the flat space limit \cite{Penedones:2010ue} is the 8d worldvolume of the sevenbranes. We can thus in principle derive stringy corrections to the sevenbrane's worldvolume theory in these F-theory compactifications by computing $1/N$ corrections to the CFT correlators. 

This program was carried out for the $D_4$ theory in \cite{Behan:2023fqq}. In this case $\tau_s$ can take any complex value, and so includes a weakly coupled regime. The dual theory is a $USp(2N)$ gauge theory with $SO(8)$ flavor symmetry and an $SL(2,\mathbb{Z})$ duality group that acts on the complexified gauge coupling $\tau\sim\tau_s$. The constraints of the analytic bootstrap \cite{Rastelli:2017udc,Alday:2021odx,Alday:2021ajh} fix the holographic correlator $M$ to take the following form 
\es{Mintro}{
M=\frac{1}{N}M_{F^2}+\frac{1}{N^2}\big[M_R+M_{F^2|F^2}+b_{F^4}M_{F^4}\big]+\frac{\log N}{N^2}b_{\log}M_{\log}+O(N^{-3})\,,
}
where we suppressed the flavor indices for simplicity, and each term is denoted by its corresponding Witten diagram in the bulk $AdS_5\times S^3$ effective theory. For instance, $M_{F^2}$ is the tree-level gluon exchange term, $M_{R}$ is the tree-level graviton exchange term, $M_{F^2|F^2}$ is the one-loop gluon term, $M_{F^4}$ is the first stringy higher-derivative correction  $F^4$ to SYM, and $M_{\log}\sim M_{F^4}$ is a contact term due to the logarithmic divergence at one loop that is regulated by string theory. The analytic bootstrap fixes all these terms at this order in $1/N$ up to the coefficients $b_{F^4}$ and $b_{\log}$, which depend on $\tau$ and require additional physical input. These coefficients were fixed in \cite{Behan:2023fqq} by considering the relation \cite{Chester:2022sqb} between integrals of the holographic correlator and the mass deformed $S^4$ free energy $F(m)$,\footnote{Similar relations have been derived for 4d $\mathcal{N}=4$ \cite{Binder:2019jwn,Chester:2020dja} and 3d $\mathcal{N}=6$ CFTs \cite{Binder:2018yvd,Binder:2019mpb}, which were used to fix $1/N$ corrections in 4d $\mathcal{N}=4$ SYM \cite{Binder:2019jwn,Chester:2020dja,Chester:2019pvm,Chester:2019jas,Chester:2020vyz,Alday:2023pet}, 3d $\mathcal{N}\geq6$ ABJM \cite{Binder:2018yvd,Binder:2019mpb,Chester:2018aca,Alday:2021ymb,Alday:2022rly,Binder:2021cif}, and another 3d $\mathcal{N}=4$ holographic theory \cite{Chester:2023qwo}. See also \cite{Pufu:2023vwo} for related constraints on Wilson loops.} which could be computed as a function of $N$ and $\tau$ using supersymmetric localization \cite{Pestun:2007rz,Pestun:2016zxk,Beccaria:2022kxy,Beccaria:2021ism}.  In the flat space limit, the $\tau$-independent value of $b_{\log}$ matched the logarithmic divergence of one-loop gluon scattering and tree-level graviton scattering off sevenbranes in ten flat dimensions, while the $\tau$-dependent value of $b_{F^4}$ matched\footnote{In fact, $b_{F^4}$ was only fixed up to a $\tau$-independent constant, because the explicit expression for $M_{R}$ is not yet known.}  the flat space $F^4$ correction as previously computed using duality to the heterotic string \cite{Bachas:1997mc,Gutperle:1999xu,Bachas:1997xn,Foerger:1998kw,Bianchi:1998vq,Gava:1999ky,Kiritsis:2000zi,Lerche:1998gz,Lerche:1998nx,Billo:2009gc,Billo:2010mg}.

The other theories in \eqref{sing} have fixed values of $\tau_s$, and are dual to rank $N$ Argyres-Douglas (AD) \cite{Argyres:1995jj,Argyres:1995xn} type CFTs for $A_1,A_2$, and Minahan-Nemeschansky (MN) \cite{Minahan:1996fg,Minahan:1996cj} type CFTs for $E_6,E_7,E_8$. These CFTs have no weakly coupled Lagrangian description.\footnote{They can be considered as the infinite Coulomb branch limit of certain gauge theories \cite{Argyres:1995jj,Argyres:1995xn,Minahan:1996cj}.} The analytic bootstrap does not require a Lagrangian, so the holographic correlator takes the same form \eqref{Mintro} as for the $D_4$ theory, except with different flavor symmetries. However, localization does require a Lagrangian, so it would seem we cannot fix the coefficients $b_{\log}$ and $b_{F^4}$ as we did for the $D_4$ theory. 

Recently, it was suggested in \cite{Russo:2014nka,Russo:2019ipg,Bissi:2021rei,Fucito:2023plp,Fucito:2023txg,Grassi:2019txd} that the sphere free energy $F(m)$ can be derived even for non-Lagrangian theories using the Seiberg-Witten curve \cite{Seiberg:1994ai,Seiberg:1994aj}, which is known even for these theories \cite{Douglas:1996js,Argyres:2002xc}. In particular, consider the free energy $F_{\epsilon_1,\epsilon_2}(m)$ on the sphere deformed by the two so-called equivariant parameters $\epsilon_1$ and $\epsilon_2$ \cite{Nekrasov:2002qd}. The round sphere partition function corresponds to setting $\epsilon_{1,2}=1$. In the $\epsilon_1,\epsilon_2\to0$ limit, we instead find the prepotential of the theory \cite{Nekrasov:2002qd}, which can also be computed from the Seiberg-Witten curve. Small $\epsilon_1,\epsilon_2$ corrections can then be computed systematically from the Seiberg-Witten curve using the holomorphic anomaly equations \cite{Bershadsky:1993ta,Huang:2006si,Huang:2009md,Krefl:2010fm,hkk11,Huang:2013eja,Codesido:2017dns,Fischbach:2018yiu}.
In \cite{Bissi:2021rei}, the sphere free energy $F$ was appoximated by keeping just the first small $\epsilon_{1,2}$ correction and then extrapolating to $\epsilon_{1,2}=1$. Certain protected OPE coefficients can be computed by taking $\tau$ derivatives of $F$ \cite{Baggio:2014sna,Baggio:2015vxa,Baggio:2014ioa}, and \cite{Bissi:2021rei} showed that these OPE coefficients as given by their approximated $F$ matched independently computed bounds from the 4d $\mathcal{N}=2$ numerical bootstrap \cite{Beem:2014zpa}. Unfortunately, it is difficult to include further small $\epsilon_{1,2}$ corrections, as the matrix model expressions for the OPE coefficients become divergent. As we will see however, there are observable quantities in \eqref{Mintro} which do not face this problem.

In this paper, we apply this Seiberg-Witten strategy to compute $F(m)$ for the F-theoretic CFTs discussed above in a large $N$ expansion. As in \cite{Bissi:2021rei}, we find that $F(m)$ is given by a matrix model whose potential includes fractional powers of the matrix eigenvalues, unlike the matrix models that appear in standard localization calculations, including for the Lagrangian $D_4$ theory discussed above. We compute the $\log N$ term that appears in the large $N$ expansion of the matrix model expression for $F(m)$, and use this to fix the $b_{\log}$ coefficient in the holographic correlator \eqref{Mintro}. We then take the flat space limit of this $AdS_5\times S^3$ holographic correlator, and show that our expression for $b_{\log}$ matches the logarithmic divergence of one-loop gluon scattering and tree-level graviton scattering off the respective configuration of sevenbranes in flat space.

The rest of this paper is organized as follows. In Section \ref{sec:theories} we discuss the theories we consider in more detail, including their holographic descriptions. In Section \ref{sec:localization} we discuss how to compute the large $N$ expansion of the sphere free energies for these theories using their Seiberg-Witten curves. In Section \ref{sec:mellin}, we use these expressions to compute the coefficient $b_{\log}$ that appears in the large $N$ expansion of the holographic correlator, and compare it to the flat space limit. We conclude in Section \ref{sec:conclusion} with a review of our results and a discussion of future directions. Technical details of the calculations are given in the various Appendices.

\section{Theories of interest and setup}\label{sec:theories}

In this work we study a set of 4d $\cN=2$ SCFTs which admit a holographic description by means of a large number $N$ of D3 branes probing certain sevenbrane singularities in F-theory \cite{Vafa:1996xn}. This setup was first explored in \cite{Sen:1996vd,Banks:1996nj,Aharony:1996en,Douglas:1996js}, where the allowed F-theory singularities were classified and the dual field theories understood. The type IIB supergravity background dual to these SCFTs was then worked out in \cite{Fayyazuddin:1998fb,Aharony:1998xz}. In this section we review some basic properties of these theories that will be relevant in the following, both from the gravity/F-theory and from the field theory perspective. We then focus on a specific observable, which will play a major role in Section \ref{sec:mellin}: the four-point function of the moment map, dual to a gluon scattering amplitude in AdS$_5\times S^3$ \cite{Alday:2021odx}. We review the superconformal kinematics and describe how, for generic 4d $\cN=2$ SCFTs, a certain integral of this correlator is related to the mass deformed sphere free energy of the theory, as shown in \cite{Chester:2022sqb}.

\subsection{The F-theory description}

We consider F-theory on a K3 singularity, which we probe with a large number $N$ of D3 branes where the dual 4d $\cN=2$ SCFTs live. Conformal invariance requires a constant axio-dilaton $\tau_s=C_0+i\,e^{-\phi}$ and this gives rise to a finite list of allowed singularities, all of which are of ADE type and are listed in \eqref{sing}. From a type IIB perspective, this corresponds to a compactification on $\mP^1$ with 24 sevenbranes. Subsets of these 24 branes generate different singularities among those in \eqref{sing}, each corresponding to the gauge groups of the 8d $\cN=1$ SYM theories living on that subset of the 24 sevenbranes. When the D3 branes probe one of these singularities (with the other sevenbranes sent to infinity), their worldvolume theory is a 4d $\cN=2$ SCFT and the Lie algebras in \eqref{sing} are associated with the flavor group $G_F$ of the SCFT. The K3 surface in the F-theory description is described by an elliptic curve
\es{elliptic_general}{
y^2=4x^3-g_2(z)\,x-g_3(z)\,,
}
where $z$ is a complex coordinate on $\mP^1$ and $g_2,\,g_3$ are polynomials in $z$. The position of the sevenbranes is determined by the vanishing of the discriminant
\es{}{
\discr(z)=16\,[g_2^3(z)-27\,g_3^2(z)]\,,
}
while the value of the axio-dilaton $\tau_s$ can be found solving $j(\tau_s)=4(24g_2)^3/\discr$. The singularities listed in \eqref{sing} correspond to the only cases allowing for constant $\tau_s$, and the corresponding values are listed in Table \ref{tab:parameters}.
\begin{table}[h!]
\centering
\begin{tabular}{|c || c | c | c|c|c|c|} 
\hline
$\mathfrak{g}$ & $A_1$ & $A_2$ & $D_4$ & $E_6$ & $E_7$ & $E_8$ \\ 
\hline
\hline
$h^{\vee}$ & 2 & 3  & 6 &  12 & 18 & 30 \\
\hline
$\nu$ & $\tfrac{1}{2}$ & $\tfrac{2}{3}$ & $1$ & $\tfrac{4}{3}$ & $\tfrac{3}{2}$ & $\tfrac{5}{3}$ \\
\hline
$n_7$ & 3 & 4 & 6 & 8 & 9 & 10\\
\hline
$\Delta$ & $\tfrac{4}{3}$ & $\tfrac{3}{2}$ & $2$ & $3$ & $4$ & $6$ \\
\hline
$\tau_s$ & $i$ & $\texp$ & any & $\texp$ & $i$ & $\texp$\\
\hline
\end{tabular}
\caption{Values of the interesting parameters for the 4d $\mathcal{N}=2$ SCFTs that we study in this paper.}
\label{tab:parameters}
\end{table}
We note that all theories arise as strongly coupled fixed points, and as such do not admit a Lagrangian description, with the exception of the $D_4$ theory. That case admits a description in terms of a superconformal gauge theory with $USp(2N)$ gauge group, with eight fundamental half-hypermultiplets (transforming in the fundamental of the flavor group $SO(8)$) and one antisymmetric hypermultiplet (singlet of $SO(8)$), and has been studied from a localization perspective in \cite{Beccaria:2021ism,Beccaria:2022kxy,Behan:2023fqq}. 

The field theories described by the F-theory singularities \eqref{sing} have been studied by several authors and a few different parameters have been introduced to identify each individual theory. We review them here for the reader's convenience, see Table \ref{tab:parameters}, and we would like to highlight that they all admit a simple expression in terms of the dual Coxeter number of the associated Lie algebras in \eqref{sing}. This is interesting because all such Lie algebras belong to the so-called Deligne's series \cite{Deligne_1996} (see also \cite{CohenMan_1996}): a set of Lie algebras for which the $n$-fold tensor product of the adjoint representation can be decomposed into irreps in a uniform way, with compact expressions for their dimensions in terms of the dual Coxeter number $\h$ -- see Appendix \ref{app:group} for more details. The first interesting parameter that we would like to discuss can be naturally introduced alongside the supergravity description of the theories in type IIB \cite{Fayyazuddin:1998fb,Aharony:1998xz}. The ten-dimensional metric is the same as the usual AdS$_5\times S^5$, with the difference that the internal space $S^5$ is replaced by $X^5$, with metric
\es{metricX5}{
ds^2(X_5)=d \theta^2+\sin^2\theta\,d \varphi^2+\cos^2\theta\,ds^2(S^3)\,,
}
where $S^3$ is a round three-sphere, $\theta \in [0,\tfrac{\pi}{2})$ and $\varphi$ is periodic with \cite{Aharony:1998xz,Fayyazuddin:1998fb}\footnote{Here we parametrize the angular deficit with $\nu$ following \cite{Alday:2021odx}, but this was originally called $\alpha$ in \cite{Aharony:1998xz}.}
\es{deficit}{
\varphi\sim \varphi+2\pi(1-\nu/2)\,, \quad
\nu=\frac{2\h}{\h+6}\,.
}
Note that \eqref{metricX5} arises from placing D3 branes on $\R^{1,3}\times \widetilde{\R}^6$, where \cite{Aharony:1998xz}
\es{}{
ds^2(\widetilde{\R}^6)=d r^2+r^2\,ds^2(X_5)=\frac{|d z|^2}{|z|^{\nu}}+ds^2(\R^4)\,,
}
with $z$ parametrizing the $\mP^1$ transverse to the sevenbranes. This coordinate, which constitutes the base of the elliptic fibration \eqref{elliptic_general}, acquires a deficit angle as in \eqref{deficit}. For $\nu=0$ the metric \eqref{metricX5} is a round $S^5$ and this case reduces to 4d $\mathcal{N}=4$ SYM. We also note that the isometries of the metric \eqref{metricX5} are $U(1)\times SO(4)$, the former factor coming from the Killing vector $\frac{\partial}{\partial\varphi}$ and the latter being the isometry of $S^3$. The $U(1)$ is dual to the $U(1)_R$ part of the R-symmetry of the dual 4d $\cN=2$ SCFT, while writing $SO(4)=SU(2)_L\times SU(2)_R$ we can identify $SU(2)_R$ as the non-abelian factor in the R-symmetry group, and $SU(2)_L$ as a global symmetry (additional to the flavor symmetry $G_F$ associated with the algebras in \eqref{sing}).

Another interesting parameter that is sometimes used to describe these theories is the conformal dimension $\Delta$ of the lightest Coulomb branch operator in the SCFTs, which is given by
\es{delta_from_h}{
\Delta=\frac{\h+6}{6}\,.
}
We note that in the cases with $D$- and $E$-type flavor Lie group, the total space of the elliptic fibration is an orbifold $T^4/\Z_{\Delta}$. The central charges of the dual SCFTs were first computed in \cite{Aharony:2007dj}, where they were conveniently expressed in terms of $\Delta$ (as well as the number $N$ of D3-branes) as
\es{centralcharges}{
a=\frac{\Delta}{4}\,N^2+\frac{\Delta-1}{2}\,N-\frac{1}{24}\,,\quad
c=\frac{\Delta}{4}\,N^2+\frac{3(\Delta-1)}{4}\,N-\frac{1}{12}\,,\quad
k=2\,\Delta\,N\,,
}
where $k$ is the flavor central charge for the global symmetry $G_F$ -- see \cite{Aharony:2007dj} for conventions. From now on we will express all results in terms of $\Delta$.

The full solution in type IIB supergravity is then given by
\es{10dsol}{
ds^2_{10}=L^2\,\left[ds^2(\text{AdS}_5)+ds^2(X_5)\right]\,,\\
F_5=4L^4\,(1+\star)\,\text{vol}(X_5)\,,
}
where $L$ is the AdS radius, $F_5$ is the five-form Ramond-Ramond flux and $\star$ denotes the ten-dimensional Hodge dual. The dilaton $\phi$ is constant, with $g_s\equiv e^{\phi}$. The flux quantization condition
\es{}{
\frac{1}{(2\pi\ell_s)^4g_s}\int_{X_5}F_5=N\,,
}
which defines the number $N$ of D3 branes, fixes
\es{dictionary}{
\left(L / \ell_s\right)^4=4 \pi\Delta g_s N\,.
}

Finally, it is also interesting to review how the singularities listed in \eqref{sing} are related to each other. Recall that type IIB string theory admits different types of (mutually non-local) sevenbranes: one typically refers to branes where $[p,q]$ strings can end as $[p,q]$ sevenbranes \cite{Witten:1995im}. As shown in \cite{Johansen:1996am,Gaberdiel:1997ud}, specific arrangements of mutually non-local branes lead to $ADE$ singularities on the transverse $\mP^1$, or equivalently $ADE$ gauge groups for the SYM theory living on the sevenbrane worldvolume. Specifically, to generate such singularities it is enough to consider $[1,0]$, $[1,-1]$ and $[1,1]$ sevenbranes, denoted as $\mathbf{A}$-, $\mathbf{B}$- and $\mathbf{C}$-branes, respectively.\footnote{We follow the conventions of \cite{Noguchi:1999xq}.} Then, the $ADE$ Lie algebras are obtained from the following sets of parallel coalescent sevenbranes
\es{}{
A_n:\,\,\, \mathbf{A}^{n+1}\mathbf{C}\,,\quad
D_n:\,\,\, \mathbf{A}^{n}\mathbf{B}\mathbf{C}\,,\quad
E_n:\,\,\,\mathbf{A}^{n-1}\mathbf{B}\mathbf{C}^2\,.
}
It follows that the singularities in \eqref{sing} are related to each other by the subsequent removal of a specific type of sevenbrane \cite{Noguchi:1999xq}
\es{7branesflow}{
E_8 
\xrightarrow{\,\,\,\mathbf{A}\,\,\,} E_7
\xrightarrow{\,\,\,\mathbf{A}\,\,\,} E_6
\xrightarrow{\,\,\mathbf{A},\mathbf{C}\,\,} D_4
\xrightarrow{\,\,\mathbf{A},\mathbf{B}\,\,} A_2
\xrightarrow{\,\,\,\mathbf{A}\,\,\,} A_1\,,
}
where the label above the arrows denotes which types of sevenbranes are sent to infinity to generate the RG flow between the various theories. Note that the flows to and from the $D_4$ theory are obtained decoupling mutually non-local sevenbranes simultaneously. We can also introduce a parameter $n_7$ which counts the number of distinct sevenbranes used in the construction for each theory, which turns out to be proportional to the parameter $\nu$ introduced in \eqref{deficit} via\footnote{The total number of sevenbranes is always 24, as required by topological consistency conditions. See \cite{Gaberdiel:1997ud} for a description of how these branes are grouped in the various cases.}
\es{}{
n_7=6\nu=\frac{12\h}{\h+6}\,.
} 

\subsection{The Seiberg-Witten description}

The theories considered in this paper are 4d $\cN=2$ SCFTs and as such they admit a description in terms of Seiberg-Witten (SW) theory \cite{Seiberg:1994ai,Seiberg:1994aj}. Let us first consider the rank one version of the theories listed in \eqref{sing}, corresponding to probing the F-theory singularities described above with a single D3-brane, as the corresponding SW curves contain all the information that we are going to need in the rest of this paper. Moreover, the $N=1$ version of these theories is also more familiar and well studied. The rank one version of the $A_1$ and $A_2$ theories are Argyres-Douglas theories \cite{Argyres:1995jj,Argyres:1995xn}, while the $E_n$ theories were first discovered by Minahan and Nemeschansky \cite{Minahan:1996fg,Minahan:1996cj}. Finally, since $USp(2)\simeq SU(2)$ and the antisymmetric multiplet decouples in this limit, the $D_4$ theory for $N=1$ is simply $SU(2)$ SQCD.

It is interesting to consider a deformation of the SCFTs described above away from the conformal point, which can be achieved by the inclusion of certain mass parameters. The inclusion of a mass deformation is dual, in the F-theory picture, to moving the sevenbranes apart from each other, and as discussed in \eqref{7branesflow} the various theories are related to each other by moving one or more of the sevenbranes to infinity, thus progressively lowering the rank of the associated Lie algebras. The inclusion of mass deformation parameters is then instrumental to the study of the renormalization group (RG) flows \eqref{7branesflow} from the field theory perspective, as first observed in \cite{Minahan:1996fg,Minahan:1996cj}. Generically, one can deform the theories by $\rank$ mass parameters, where $\rank$ is the rank of the Lie algebras in \eqref{sing}, and after turning on the mass deformation, one can perform a suitable scaling limit in the respective SW curves which involves sending one (or two, in the flows $E_6\to D_4$ and $D_4\to A_2$) of the masses to infinity. As we shall see, it is crucial for the rest of this paper that, thanks to the observations in \cite{Noguchi:1999xq}, it is possible to perform this RG flow while keeping the same normalization for the mass parameters in all theories.  

While, as we just reviewed, the most generic mass deformation is well understood, for our purposes it is enough to focus on a simpler setup. Indeed, the main quantity that we are going to consider here is the fourth mass derivative of the free energy, evaluated for zero mass: this allows us to consider deformations that are at most quartic in the mass parameters. Moreover, all Lie algebras in \eqref{sing} have exactly one quadratic Casimir and one quartic Casimir -- which is simply the square of the former -- with the exception of $D_4$, which we discuss later. Hence, it is enough to consider a single mass parameter $m$ for all theories, in terms of which the quadratic Casimir is $m^2$ and the quartic one is $m^4$ (up to normalizations). Doing so corresponds to turning on a mass deformation for the Cartan of the common $SU(2)$ subgroup to all flavor groups. The most generic deformation for the $D_4$ theory that is quartic in mass contains more information than this, and it was considered from this perspective in \cite{Behan:2023fqq}.

Another caveat concerns the normalization of the various terms entering the SW curves. In the following it will be particularly important to have a uniform definition of the mass parameter $m$ discussed above, which does not change along the RG flows \eqref{7branesflow}: for this reason, we find it particularly useful to follow the presentation of \cite{Noguchi:1999xq}, which gives a prescription allowing precisely for that. Another advantage of \cite{Noguchi:1999xq} is to clarify the confusion of \cite{Minahan:1996fg,Minahan:1996cj} regarding whether or not the representation of the gauge group chosen for the SW differential $\lambdaSW$ affects the physics, answering this question negatively and giving a prescription to normalize $\lambdaSW$ in a uniform way for all groups. Here we adapt the results of \cite{Noguchi:1999xq} to our conventions, in the following sense. The SW curves and differentials of \cite{Noguchi:1999xq} are given as
\es{}{
\tilde{y}^2=\tilde{x}^3+f(\tilde{z})\,\tilde{x}+g(\tilde{z})\,,\quad
\frac{\de \lambdaSW}{\de \tilde{z}}=\kappa_{\mathfrak{g}}\,\frac{d \tilde{x}}{\tilde{y}}+d(\ldots)\,,
}
where $d(\ldots)$ denotes a total derivative term and the second equation fixes the scale for the Coulomb branch parameter $\tilde{z}$ (the equivalent of the $\mP^1$ coordinate $z$ in \eqref{elliptic_general}), where $\kappa_{\mathfrak{g}}$ is a constant which depends on the choice of Lie algebra $\mathfrak{g}$ and can be read off from the results in \cite{Noguchi:1999xq}. We perform the following rescalings
\es{}{
\tilde{y}\to \sqrt{2}y\,,\quad \tilde{x}\to 2 x\,,\quad
\tilde{z}\to \frac{1}{\sqrt{2}\pi\,\kappa_{\mathfrak{g}}} u\,,
}
which bring the curve and differential in the form
\es{SWgeneral}{
y^2=4x^3-g_2(u)\,x-g_3(u)\,,\quad
\frac{\de \lambdaSW}{\de u}=\frac{1}{\pi}\,\frac{d x}{y}+d(\ldots)\,,
}
in terms of the new Coulomb branch parameter $u$, which is now normalized uniformly for all groups. Finally, \cite{Noguchi:1999xq} introduces mass parameters $\phi_i$, $i=1,\ldots,\rank$ and we set $\phi_i=0$ for $i>1$ and $\phi_1=m$, where the choice of relative normalization between $\phi_1$ and $m$ (simply one, in this case) is crucial, and is chosen in such a way that $m$ is exactly the mass parameter of the $D_4$ theory obtained setting $\mu_1=m$ and $\mu_2=\mu_3=\mu_4=0$ in \cite{Chester:2022sqb,Behan:2023fqq}. This particular normalization is going to be crucial when we consider the integrated correlator constraint. We also remark again that we only need to fix the ratio between $\phi_1$ and $m$ in one of the theories (the $D_4$ one, in this case), thanks to the prescription of \cite{Noguchi:1999xq} that fixes the relative normalization between mass parameters and SW differentials. With these caveats, the mass deformed SW curves we are interested in have the form \eqref{SWgeneral}, where the functions $g_2(u)$ and $g_3(u)$ are given in Table \ref{tab:SWcruves_g23}.
\begin{table}[h!]
\centering
\begin{tabular}{|c || c | c |c|c|c|} 
\hline
$\mathfrak{g}$ & $A_1$ & $A_2$ &  $E_6$ & $E_7$ & $E_8$ \\ 
\hline
\hline
$g_2(u)$ & $8u$ & $4m^2$  & $2m^2u^2$ & $8u^3$ & $2m^2u^3$\\
\hline
$g_3(u)$ & $2m^2$ & $8u^2$  & $8u^4$ & $\tfrac{2}{3}m^2u^4$ & $16u^5$\\
\hline
\end{tabular}
\caption{Functions $g_2(u)$ and $g_3(u)$ for the SW curves of the various SCFTs, using the representation \eqref{elliptic_general}.}
\label{tab:SWcruves_g23}
\end{table}
The SW curve of the $D_4$ theory is more complicated since $g_2$ and $g_3$ are also functions of the complexified gauge coupling $\tau$. The expression of the curve as a function of $u$, $\tau$ and four independent mass parameters $m_1,\ldots,m_4$ can be found in \cite{Seiberg:1994aj}. Here we present that curve in the limit where only one mass is non-vanishing, after adapting it to our conventions. We can start from eq. (2.10) of \cite{Noguchi:1999xq} and with a suitable change of variables we obtain\footnote{To obtain this, start from eq. (2.10) of \cite{Noguchi:1999xq} and set $m_1=m$ and $m_2=m_3=m_4=0$, while changing coordinates with
\es{}{
Y\to \sqrt{2}y\,,\quad
X\to 2x+\frac{m^2}{6}(\alpha-\beta)^2+\frac{4}{3}(\alpha+\beta)u\,,\quad
Z\to  4u\,.
}
}
\es{g23_D4}{
g_2=&\frac{m^4}{12}(\alpha-\beta)^2(\alpha^2+\alpha\beta+\beta^2)+\frac{4m^2}{3}(\alpha-\beta)^2(\alpha+\beta)u+\frac{16}{3}(\alpha^2-\alpha\beta+\beta^2)u^2\,,\\
g_3=&\frac{m^6}{864}(\alpha-\beta)^4(2\alpha+\beta)(\alpha+2\beta)+\frac{m^4}{18}(\alpha-\beta)^2(\alpha+\beta)(2\alpha^2-\alpha\beta+2\beta^2)u\\
&+\frac{4m^2}{9}(\alpha-\beta)^2(2\alpha^2+\alpha\beta+2\beta^2)u^2+\frac{32}{27}(\alpha-2\beta)(2\alpha-\beta)(\alpha+\beta)u^3\,,
}
where following \cite{Noguchi:1999xq} we have used
\es{}{
\alpha=-\theta^4_3(\tau)\,,\quad
\beta=-\theta^4_4(\tau)\,,
}
which are elliptic Jacobi theta functions of the complexified gauge coupling $\tau$ of the $D_4$ theory.

To conclude, we add that while here we have discussed the SW curves for the rank one version of the theories of interest, the results can be trivially generalized to general rank. The relation between the two is particularly simple for the prepotential, where in the cases of interest the prepotential for the rank $N$ theory is simply given by the sum of $N$ copies of the rank one prepotential, see \eqref{f0-known} and the discussion of Section \ref{sec:localization} for more details.

\subsection{The four-point function of the moment map}

The main observable that we are going to consider is the four-point function of the moment map operator $\phi^A$ in the 4d $\mathcal{N}=2$ SCFTs listed in \eqref{sing}. This is the superconformal primary operator of the flavor current multiplet and it is a Lorentz scalar of conformal dimension 2 transforming in the adjoint of the $SU(2)_R$ part of the R-symmetry, while being a scalar under the $U(1)_R$ part. It is also in the adjoint representation of the flavor group $G_F$ associated with the Lie algebras in \eqref{sing} in each case and $A=1,\ldots,\text{dim}(G_F)$, while it is a singlet under the additional $SU(2)_L$ factor of the global symmetry group, which we discussed below \eqref{metricX5}. The flavor current multiplet is dual to the gluon supermultiplet in AdS$_5$, in the dual gravitational description. As first discussed in \cite{Alday:2021odx}, its four-point function is then dual to the scattering of open strings in AdS$_5\times S^3$, which is the backreacted geometry on the worldvolume of the sevenbranes, where $S^3$ is the singular locus of the metric \eqref{metricX5}.

In terms of spinor polarizations $y$ for $SU(2)_R$, invariance under the bosonic part of the superconformal algebra and under global symmetries require the four-point function of $\phi^A$ to be of the form
\es{momentmapfourpoint}{
\langle \phi^A(x_1;y_1)\phi^B(x_2;y_2)\phi^C(x_3;y_3)\phi^D(x_4;y_4)\rangle=\frac{\langle y_1,y_2\rangle^2\langle y_3,y_4\rangle^2}{x_{12}^4x_{34}^4}G^{ABCD}(U,V;w)\,,
}
in terms of the cross ratios
\es{crossratios}{
U=\frac{x_{12}^2x_{34}^2}{x_{13}^2x_{24}^2}\,,\quad
V=\frac{x_{14}^2x_{23}^2}{x_{13}^2x_{24}^2}\,,\quad
w=\frac{\langle y_1,y_2\rangle \langle y_3,y_4\rangle}{\langle y_1,y_3\rangle \langle y_2,y_4\rangle}\,,
}
with $x_{12}=x_1-x_2$ the four-dimensional distances and $\langle y_1,y_2\rangle=y_1^iy_2^j\epsilon_{ij}$ the $SU(2)_R$-invariant distances, with $i,j=1,2$. The function $G^{ABCD}(U,V;w)$ can be expanded as
\es{Girreps}{
G^{ABCD}(U,V;w)=\sum_{\mathbf{R} \in \mathfrak{g}\otimes \mathfrak{g}}G_\mathbf{R}(U,V;w)\,P_\mathbf{R}^{ABCD}\,,
}
in terms of projectors $P_\mathbf{R}^{ABCD}$. In Appendix \ref{app:group} we discuss the decomposition of $\mathfrak{g}\otimes \mathfrak{g}$ into irreducible representations for all the Lie algebras in \eqref{sing}, where we denote with $\mathfrak{g}$ the adjoint representation of a Lie algebra $\mathfrak{g}$, and provide a basis of tensor structures in terms of which we give the projectors $P_\mathbf{R}^{ABCD}$ in a uniform way, with coefficients that depend uniquely on the dual Coxeter number $\h$.

The flavor current multiplet is one-half BPS and its four-point function is therefore subject to further constraints by superconformal invariance, known as the superconformal Ward identities \cite{Dolan:2001tt}
\es{}{
\left. \left(z\,\partial_z+w\,\partial_w\right)G_\mathbf{R}(U,V;w)\right|_{w=z}=0=\left. \left(\bar{z}\,\partial_{\bar{z}}+w\,\partial_w\right)G_\mathbf{R}(U,V;w)\right|_{w=\bar{z}}\,,
}
which can be formally solved by writing, for each representation $\mathbf{R}$,
\es{reducedcorrelator}{
G_\mathbf{R}(U,V;w)=\frac{z(w-\bar{z})f_\mathbf{R}(\bar{z})-\bar{z}(w-z)f_\mathbf{R}(z)}{w(z-\bar{z})}+\left(1-\frac{z}{w}\right)\left(1-\frac{\bar{z}}{w}\right)\mathcal{G}_\mathbf{R}(U,V)\,,
}
where $U=z\bar{z}$, $V=(1-z)(1-\bar{z})$ and we introduced the reduced correlator $\mathcal{G}$ as well as the holomorphic correlator $f$. The latter is a protected quantity that can be computed from the chiral algebra associated to any 4d $\cN=2$ SCFT and reads \cite{Beem:2013sza}
\es{}{
f^{ABCD}(z)=\delta^{AB}\delta^{CD}+z^2\delta^{AC}\delta^{BD}+\frac{z^2}{(1-z)^2}\delta^{AD}\delta^{BC}+\frac{2z}{k}f^{ACE}f^{BDE}+\frac{2z}{k(z-1)}f^{ADE}f^{BCE}\,,
}
in terms of the flavor central charge $k$ introduced in \eqref{centralcharges} and the structure constants $f^{ABC}$ associated with the Lie algebras in \eqref{sing} -- see Appendix \ref{app:group} for our conventions.

\subsection{Integrated correlator constraints}

On top of the constraints on the moment map four-point function that are dictated by superconformal invariance, it was shown in \cite{Chester:2022sqb} (building on the results of \cite{Binder:2018yvd,Binder:2019mpb,Binder:2019jwn,Chester:2020dja}) that in any 4d $\cN=2$ SCFT a certain integral of this correlator can be related to the mass deformed sphere free energy $F(m)$, which can be computed exactly using supersymmetric localization \cite{Pestun:2007rz}. 

More precisely, let us consider the integral
\es{integral}{
\left.I[\mathcal{G}] \equiv \frac{1}{\pi} \int d R d \theta R^3 \sin ^2 \theta \frac{\bar{D}_{1,1,1,1}(U, V) \mathcal{G}(U, V)}{U^2}\right|_{\substack{U=1+R^2-2 R \cos \theta \\ V=R^2}}\,,
}
where 
\es{}{
\bar{D}_{1,1,1,1}(U, V)=\frac{1}{z-\bar{z}}\left(\log (z \bar{z}) \log \frac{1-z}{1-\bar{z}}+2 \operatorname{Li}(z)-2 \operatorname{Li}(\bar{z})\right)\,.
}
It was shown in \cite{Chester:2022sqb} that this is related to the fourth mass derivative of the sphere free energy, with the mass parameters set to zero after taking the derivatives. Hence, the number of constraints implied by this relation equals the number of quartic Casimirs of the flavor group. In the case at hand, we note that all groups in \eqref{sing} have only one quartic Casimir, which is the square of the quadratic Casimir, so that as described earlier in this section it is sufficient to consider a single mass parameter $m$, in terms of which the quadratic and quartic Casimirs are $m^2$ and $m^4$, respectively. The only exception is $D_4$, which admits three quartic Casimirs and was considered from this perspective in \cite{Behan:2023fqq}. However, this will not be relevant for the quantities that we consider here and indeed there is a sense in which $D_4$ has analogous properties to the other Lie algebras in \eqref{sing}, as we explain in Appendix \ref{app:group}. The restriction to only one mass is equivalent to taking all the flavor indices to be identical in the four-point function, say $A=B=C=D=1$ for definiteness. The constraint of \cite{Chester:2022sqb} then reads
\es{integrated_constraint}{
\left. -\partial^4_m F(m)\right|_{m=0}=k^2\,I[\mathcal{G}^{1111}_{\text{conn}}]\,,
}
where we have introduced the connected part $\mathcal{G}^{ABCD}_{\text{conn}}$ of the reduced correlator $\mathcal{G}^{ABCD}$, which is defined via
\es{}{
\mathcal{G}^{ABCD}_{\text{conn}}=\mathcal{G}^{ABCD}-\left[\delta^{AB}\delta^{CD}+\delta^{AC}\delta^{BD}\,U+\delta^{AD}\delta^{BC}\,U/V^2\right]\,.
}
See Appendix \ref{app:group} for a definition of $\mathcal{G}^{1111}$ in terms of a convenient basis of tensors and in terms of irreducible representations appearing in $\mathfrak{g}\otimes \mathfrak{g}$.

We end this section with a comment on the choice of normalization for the mass parameter $m$ appearing in the SW curves listed in Table \ref{tab:SWcruves_g23}. The integrated correlator constraint \eqref{integrated_constraint} is obtained as a limit of that considered in \cite{Chester:2022sqb,Behan:2023fqq} for theories with $SO(8)$ flavor symmetry, where we have taken $\mu_1\equiv m$ and $\mu_2=\mu_3=\mu_4=0$. So the normalization of $m$ is under control in the $D_4$ theory and indeed using \eqref{integrated_constraint} we can reproduce the results of \cite{Behan:2023fqq} which can be studied in this limit. As we emphasized in the discussion of the mass deformed SW curves, we have been careful not to change the normalization of $m$ in flowing between the various theories, which allows us to phrase the constraint as in \eqref{integrated_constraint} for all cases. Note that this normalization is also the reason for the choice $\phi_1=m$ in adapting the results of \cite{Noguchi:1999xq}.

\section{Computing the sphere free energy}\label{sec:localization}

To study the four-point function dual to gluon scattering, we will use the integrated constraint \eqref{integrated_constraint} which depends on the mass deformed sphere free energy. In general the number of masses by which one can deform is the rank $\mathtt{r}$ as explained in Section \ref{sec:theories}. For our purposes however, we can use only one mass and write $F(m) \equiv -\log Z(m)$ because the $A_1$, $A_2$ and $E_n$ groups have only a single quartic Casimir. We must now scrutinize the partition function
\es{ZSW}{
Z(m) = \int d^Na \left | e^{r^2 \mathcal{F}(a, m, \Lambda, r)} \right |^2\,,
}
where $\mathcal{F}(a, m, \Lambda, r)$ is the SW prepotential and $\Lambda$ is the renormalization invariant scale which is needed for a non-conformal theory. It was explained in \cite{Russo:2014nka} that the relation between the prepotential and partition function of a gauge theory holds for a wider class of theories due to the latter's dependence on $\Lambda$. These are the 4d $\mathcal{N} = 2$ QFTs which can be reached from a gauge theory in the UV. The definition of the $\Omega$ background prepotential from \cite{Nekrasov:2002qd}, generalizing the flat space definition in \cite{Seiberg:1994ai,Seiberg:1994aj}, then goes through in this case and \eqref{ZSW} is the statement of how one uses the $\Omega$ background to recover the partition function on a squashed sphere \cite{Pestun:2007rz}. Computing $\mathcal{F}(a, m, \Lambda, r)$ exactly is a difficult problem which will not be solved here. Progress usually comes from one of two approaches. For Lagrangian theories, an equivalent formulation of the partition function is
\es{ZS4}{
Z(m) = \int [dA] |\Lambda^{\beta_1}|^{\frac{1}{2} \text{tr} A^2} |Z_\text{1-loop}(A, m)|^2 |Z_\text{inst}(A, m, \Lambda)|^2\,,
}
where the one-loop and instanton contributions generalize more readily to backgrounds preserving a supercharge through supersymmetric localization \cite{Nekrasov:2002qd,Nekrasov:2003rj,Shadchin:2005mx,Pestun:2007rz}. Here, $\beta_1$ appears in the beta function as $\beta(\tau) = -\frac{g_\text{YM}^3 \beta_1}{16 \pi^2} + O(e^{2\pi i \tau})$ and $[dA]$ is a measure which is known for any classical gauge group.\footnote{With $USp(2N)$ for example, it is given by
\es{Zmeasure}{
[dA] = \frac{1}{N!} \prod_{i = 1}^N da_i a_i^2 \prod_{j < k} (a_j^2 - a_k^2)^2\,.
}
}
Although \eqref{ZS4} still has a relation to non-Lagrangian theories which live at infinite Coulomb branch points \cite{Russo:2014nka,Russo:2019ipg}, one typically does not have enough information to take the $\Lambda \to \infty$ limit for finely tuned masses. As such, we will use a different approach which directly makes use of \eqref{ZSW} but expands the prepotential around flat space as
\es{prepot-exp-schematic}{
\mathcal{F}(a) = \mathcal{F}_0(a) + r^{-2} \mathcal{F}_1(a) + r^{-4} \mathcal{F}_2(a) + O(r^{-6})\,.
}
Computing instanton effects exactly but having to expand in $r$ complements traditional localization calculations which are exact in $r$ but only tractable at weak coupling. Fortunately, the $\log N$ contribution we are after will only require the first two terms of \eqref{prepot-exp-schematic}.\footnote{The prepotential to order $1/r^2$ was also used in the uncontrolled calculation of \cite{Bissi:2021rei}. Cutting off the expansion after two terms happened to give OPE coefficients comparable to those of the numerical bootstrap but there was no \textit{a priori} reason for why this should have been the case. By contrast, we will see that none of the terms in \eqref{prepot-exp-schematic} beyond $1/r^2$ can produce a $\log(N)$ in the large $N$ expansion \eqref{Mintro}. The truncation used in \cite{Bissi:2021rei} will therefore be controlled in our case, allowing for an exact match between the coefficient of such $\log(N)$ term in the Mellin amplitude we consider and the $\log\ell_s$ term appearing in the corresponding flat space amplitude, after taking the flat space limit. See Section \ref{sec:mellin} for more details.}

\subsection{Mass deformed prepotentials}

We will work with the prepotential as a function of the gauge invariant coordinates $u_i$ instead of the Coulomb branch eigenvalues $a_i$. The period integrals defined in \cite{Seiberg:1994ai,Seiberg:1994aj}, to be discussed shortly, provide a link between the two. The $\log N$ term in the free energy will come from the $\sum_i \log u_i$ term in the prepotential which must be quadratic in mass by dimensional analysis. On a general $\Omega$ background with radius $r = \frac{1}{\sqrt{\epsilon_1 \epsilon_2}}$ and squashing parameter $b = \sqrt{\frac{\epsilon_1}{\epsilon_2}}$, the prepotential has the expansion
\es{prepot-exp}{
\mathcal{F}(u) = \mathcal{F}_0(u) + r^{-2} \left [ \frac{1}{2} \log \det \left ( \frac{\partial u}{\partial a} \right ) + \frac{b^2 + b^{-2}}{24} \log \mathbf{\Delta}(u)\right ] + O(r^{-4})\,,
}
where $\mathbf{\Delta}(u)$ is the discriminant of the SW curve \cite{mw97,ny03,Shapere:2008zf}. Setting $b = 1$, this turns \eqref{ZSW} into
\es{ZSW-flat}{
Z(m) \approx \int d^Nu \, \mathbf{\Delta}(u)^{1/6} \left | e^{r^2 \mathcal{F}_0(u, m)} \right |^2\,,
}
where the approximation results from cutting off \eqref{prepot-exp} at $O(r^{-4})$. It was found in \cite{Douglas:1996js} that the hyperelliptic curves for mass deformed CFTs engineered from F-theory factorize into a product of elliptic curves.\footnote{For the theories obtained from the $USp(2N)$ conformal manifold, this is only true when the anti-symmetric hyper is massless. Turning on a mass for it makes the SW curve genuinely hyperelliptic and the flat space prepotential is no longer a simple sum \cite{Argyres:2002xc,acl21}.} This means that the flat space prepotential of a rank $N$ theory can be written as
\es{f0-known}{
\mathcal{F}_0(u) = \sum_{i = 1}^N \widetilde{\mathcal{F}}_0(u_i)\,,
}
where $\widetilde{\mathcal{F}}_0$ is the flat space prepotential for $N = 1$. Equivalently, each $u_i$ is a function of a single $a_i$. The discriminant $\mathbf{\Delta}(u)$ is not simply the product of the rank one discriminants $\widetilde{\mathbf{\Delta}}(u)$ due to an extra term which plays the role of a Vandermonde determinant. For the theories of interest here, \cite{Shapere:2008zf} found the expression
\es{f1-known}{
\mathbf{\Delta}(u) = \prod_{i < j} (u_i - u_j)^6 \prod_{i = 1}^N \widetilde{\mathbf{\Delta}}(u_i)\,.
}
Several higher curvature corrections to the rank one prepotentials were recently computed in \cite{Fucito:2023txg}. For any rank, \cite{Huang:2006si,Huang:2009md,hkk11,Huang:2013eja} showed that they can be fixed recursively using an anomaly equation.\footnote{Depending on the context, this is referred to as either a holomorphic or modular anomaly equation. This comes from the fact that $E_2(\tau, \bar{\tau}) = E_2(\tau) - \frac{3}{\text{Im}(\tau)}$ relates a holomorphic quasi-modular form to a non-holomorphic modular form.} 

The calculation to be done is effectively rank one so we may focus on a single copy of the elliptic curve
\es{curve-reminder}{
y^2 = 4x^3 - g_2(u)x - g_3(u)\,,
}
as given explicitly in Section \ref{sec:theories} in terms of the one mass deformation needed here. This is proportional to the first of the $\mathtt{r}$ parameters $\phi_i$ from \cite{Noguchi:1999xq} without loss of generality. With the single mass $m = \phi_1$ turned on, the flat space prepotential takes the form
\es{f0-form}{
\widetilde{\mathcal{F}}_0(u) = -\frac{c^2_\mathcal{F}}{2} u^{2/\Delta} - f_\text{log} \, m^2 \log u + O(m^4)\,,
}
and gives access to the important coefficient $f_\text{log}$. The precise value of $c^2_\mathcal{F}$ will not be important for us but its real part must be positive for \eqref{ZSW-flat} to converge. To proceed, the periods which determine the running coupling through $\tau \equiv \omega_2 / \omega_1$ are defined by
\es{periods1}{
\omega_1 = \frac{\partial a}{\partial u} = \oint_\alpha \frac{\partial \lambdaSW}{\partial u}, \quad \omega_2 = \frac{\partial a_D}{\partial u} = \oint_\beta \frac{\partial \lambdaSW}{\partial u}\,,
}
where $\alpha, \beta$ are two independent cycles and $\lambdaSW$ is the SW differential.\footnote{The differential $\lambdaSW$ is naturally associated with a representation of the flavor group but the quantity $\frac{\partial \lambdaSW}{\partial u}$ used above is independent of the representation as shown in \cite{Noguchi:1999xq}. This corrects an earlier claim made in \cite{Minahan:1996cj}.} As shown in \eqref{SWgeneral}, our convention is to take
\es{sw-diff}{
\frac{\partial \lambdaSW}{\partial u} = \frac{1}{\pi} \frac{dx}{y}\,,
}
up to an exact differential. Studies which only turn on a mass to regulate the approach to a conformal point (\textit{e.g.} \cite{Fucito:2023txg}) are able to impose \eqref{sw-diff} arbitrarily without changing the curves but we will need to be more careful. As discussed in Section \ref{sec:theories}, \cite{Noguchi:1999xq} computes curves and differentials which are consistent in the sense that masses stay fixed along the flow from one theory to the next after removing branes. We have therefore arrived at Table \ref{tab:SWcruves_g23} by taking the curves of \cite{Noguchi:1999xq} and rescaling $u$ in precisely the way that guarantees \eqref{sw-diff}. The final step of setting $\phi_1 = m$ ensures that $m$ in the $D_4$ theory (and therefore all the others) is the one that multiplies a canonically normalized moment map operator.

With this information, equations which follow from \eqref{periods1} (see \textit{e.g.} \cite{Fucito:2023txg}) are 
\es{periods2}{
g_2(u) = \frac{E_4(\tau)}{3 \omega_1(u)^4}\,, \quad g_3(u) = \frac{8 E_6(\tau)}{27 \omega_1(u)^6}\,,
}
in terms of the weight 4 and 6 Eisenstein series. Note that the values of $g_2$ and $g_3$ for the theories of interest are collected in Table \ref{tab:SWcruves_g23} and eq. \eqref{g23_D4}, including the mass deformation relevant for this paper, while we collect definitions and properties of the Eisenstein series $E_k(\tau)$ in Appendic \ref{app:modular}. By definition, the flat space prepotential has the expression
\es{int-f0}{
\widetilde{\mathcal{F}}_0(u) = -2\pi i \int^u du' \, \omega_1(u') \int^{u'} du'' \, \omega_1(u'') \tau(u'')\,.
}
Conformal points are especially easy to handle because the coupling becomes a constant independent of $u$ (either $i$ or $\texp$ as listed in Table \ref{tab:parameters}). Using the values of $E_4$ and $E_6$ at these points given in \eqref{einsenstein_fixedtau}, we can evaluate \eqref{int-f0} trivially. At a mass deformed point, one has to work a bit harder. A good strategy is to write down an ansatz which expands $\tau$ in even powers of $m$ and then fix the coefficients self-consistently using the identities \eqref{eiseinstein_derivative} for the $\tau$ derivatives of $E_4$ and $E_6$ which arise in the small mass expansion. For the $A_1$ and $E_7$ theories, we set $\tau = i + \tau^{(2)} m^2 + O(m^4)$ and rearrange \eqref{periods2} to get
\es{e6-sol}{
E_6(\tau) = g_3(u) \left [ \frac{3 E_4(\tau)}{g_2(u)} \right ]^{3/2}\,,
}
where both the functions $g_2$, $g_3$ (see Table \ref{tab:SWcruves_g23}) and the arguments of the Eisenstein series depend on $m$, and note that due to \eqref{einsenstein_fixedtau} the left-hand side vanishes for $m=0$. Expanding both sides to the lowest non-trivial order in $m$ is enough to fix $\tau^{(2)}$. Similarly, $A_2$, $E_6$ and $E_8$ tell us to set $\tau = \texp + \tau^{(2)} m^2 + O(m^4)$. In these cases, it is useful to rearrange the equation above as
\es{e4-sol}{
E_4(\tau) = \frac{g_2(u)}{3} \left [ \frac{E_6(\tau)}{g_3(u)} \right ]^{2/3},
}
with the dependence on $m$ as before, in such a way that again the left-hand side vanishes for $m=0$ due to \eqref{einsenstein_fixedtau}. The first non-trivial order in the expansion for small $m$ fixes $\tau^{(2)}$. Finally, we can consider the $D_4$ theory.\footnote{A normalization convention for the superpotential is given in \cite{Seiberg:1994aj} which directly associates their masses with canonically normalized moment maps. The relation $\phi_1 = m$ has been concluded from the fact that the curve in \cite{Seiberg:1994aj} precisely agrees with \eqref{g23_D4} in this case after our rescaling of $u$.} Here, the complexified gauge coupling is arbitrary so we set $\tau=\tau^{(0)}+\tau^{(2)}m^2+O(m^4)$. We can then consider \eqref{periods2} in either of the forms \eqref{e6-sol} or \eqref{e4-sol}, where $g_2$ and $g_3$ are given in \eqref{g23_D4}, with the caveat that we should replace $\tau$ in those expression with the mass-independent $\tau^{(0)}$, while the argument of the Eisenstein series is the mass-dependent $\tau$. Expanding this equation for small mass up to $O(m^2)$ allows to fix $\tau^{(2)}$ in this case as well, recalling the relations \eqref{E_from_theta} between Eisenstein series and theta functions as well as the Jacobi identity \eqref{jacobi_identity}. The arguments above give $\tau^{(2)}$ in all cases and we can use the result to compute the first period $\omega_1$, for which we also consider a small mass expansion of the form $\omega_1 = \omega_1^{(0)} + O(m^2)$. We are only interested in the leading order, which follows directly from \eqref{periods2}. To express our results, it is convenient to introduce two constants $c_1$ and $c_2$ via
\es{final_constants}{
\tau^{(2)} = c_1 m^2 u^{-2 \Delta^{-1}}, \quad \omega_1^{(0)} = c_2 u^{\Delta^{-1} - 1}\,,
}
whose value in the various cases we collect in Table \ref{tab:flog_info}.
\begin{table}[h!]
\centering
\begin{tabular}{|c||c|c|}
\hline
$\mathfrak{g}$& $c_1$ & $c_2$  \\
\hline
\hline
$A_1$ & $\frac{3i}{16\pi} \sqrt{6/E_4(i)}$ & $[E_4(i)/6]^{1/4}$  \\
$A_2$ & $\frac{i\sqrt{2}}{\pi} [16\sqrt{2}E_6(\texp)]^{-1/3}$ & $[E_6(\texp)/27]^{1/6}$  \\
$D_4$ & $\tfrac{i}{16\pi}$ & $\sqrt{2}$ \\
$E_6$ & $\frac{i\sqrt{2}}{\pi} [128\sqrt{2}E_6(\texp)]^{-1/3}$ & $[E_6(\texp)/27]^{1/6}$  \\
$E_7$ & $\frac{i}{16\pi} \sqrt{6/E_4(i)}$ & $[E_4(i)/6]^{1/4}$  \\
$E_8$ & $\frac{i\sqrt{2}}{\pi} [512\sqrt{2}E_6(\texp)]^{-1/3}$ & $[E_6(\texp)/54]^{1/6}$  \\
\hline
\end{tabular}
\caption{Constants which appear in the first period of the SW curve and the first massive correction to the gauge coupling according to \eqref{final_constants}. These determine the prepotential to the order shown in \eqref{f0-form}.}
\label{tab:flog_info}
\end{table}
Evaluating the integral \eqref{int-f0}, the desired coefficient in the prepotential is
\es{final_flog}{
f_\text{log} = -2\pi i \Delta c_1 c_2^2 = \frac{1}{2}\,.
}
It is worth stressing that it is the same in all six cases and that this has not been put in by hand at any stage. For $D_4$, our solution for $f_\text{log}$ is consistent with the results of \cite{Behan:2023fqq} as one can see from the free energy calculations of either the next subsection or Appendix \ref{app_usp}. For all other flavor groups, this is a new result.

\subsection{Analytically continuing the matrix model}

Our calculations so far have shown that $f_\text{log} = 1/2$ for all types of F-theory singularities that lead to a CFT. The next step is using \eqref{ZSW-flat} to convert this to a statement about the sphere free energy. Referring to \eqref{f0-known} and \eqref{f1-known}, the partition function is given by
\es{Zcanonical}{
Z(m) = \int d^N u \, \prod_{i < j} |u_i - u_j| \exp \left [ -N c^2_{\mathcal{F}} \sum_{i = 1}^N u_i^{2/\Delta} + V(u, m) \right ]\,,
}
where we have rescaled $u_i \to N^{\Delta/2} u_i$ and dropped an irrelevant overall factor. The most favorable situation occurs when $V(u, m)$ can be written as a series involving non-negative powers of $u_i$. These operators have well defined expectation values with respect to the unperturbed matrix model (which should more properly be called an eigenvalue ensemble). Since \eqref{prepot-exp} takes the form of an expansion around flat space, we are in a less favorable situation here. Additional factors of $r^{-2}$ necessarily come with negative powers of $u_i$ meaning our expression for $V(u, m)$ is most trustworthy when the eigenvalues are large instead of small. The way out will be to assume that the closed form expressions for $\left < \text{tr} \, u^{k_1} \dots \text{tr} \, u^{k_n} \right >_\mathrm{conn}$ connected correlators, which are \textit{a priori} valid for $k_i > 0$, can be analytically continued outside this domain. The validity of this assumption can be tested on the $D_4$ theory studied in \cite{Beccaria:2021ism,Beccaria:2022kxy,Behan:2023fqq} whose matrix model is known as an expansion in both large and small eigenvalues. Appendix \ref{app_usp} compares the two approaches and finds the expected agreement.

With the analytic continuation discussed above, it is clear that the $\log N$ part of the sphere free energy which is quartic in mass comes from a single correlation function and in particular we have
\es{F_loglog}{
F=-2f^2_{\log}m^4\,\left< \sum_{i = 1}^N \log(N^{\Delta / 2} u_i) \sum_{j = 1}^N \log(N^{\Delta / 2} u_j) \right>_\mathrm{conn}+O(1)\,,
}
so that we only have to compute a two-point function from which to extract the desired $\log N$ term. As it turns out, it has been known for a long time that the two-point function at large $N$ admits a universal\footnote{As discussed in \cite{ekr15}, the two-point function in a matrix model with a large $N$ limit is independent both of the power of the Vandermonde determinant and of the specific potential chosen. The expression used here is valid under the ``one-cut'' assumption, which essentially amounts to the fact that at large $N$ the eigenvalues have a distribution supported on a single interval  -- see \cite{ekr15} for details.} result in generic matrix models and it was first derived in \cite{Ambjorn:1990ji} -- see also \cite{ekr15} for a more recent exposition and a review of general matrix models. We find
\es{universal-2pt}{
\left<\sum_{i = 1}^N u_i^{2k_1} \sum_{j = 1}^N u_j^{2k_2}\right>_\mathrm{conn}=\frac{1}{\pi} \frac{\Gamma(k_1 + 1/2) \Gamma(k_2 + 1/2)}{\Gamma(k_1) \Gamma(k_2)} \frac{\nu^{2k_1 + 2k_2}}{k_1 + k_2}+O(1/N)\,,
}
where $\nu$ is the spectral edge parameter, which we introduce later but is unimportant so far as the dependence of \eqref{F_loglog} on $\log N$ is concerned. We can now compute
\begin{align}
\left < \sum_{i = 1}^N \log(N^{\Delta / 2} u_i) \sum_{j = 1}^N \log(N^{\Delta / 2} u_j) \right >_\mathrm{conn} &= \frac{1}{4} \lim_{k_1, k_2 \to 0} \partial_{k_1} \partial_{k_2} N^{\Delta (k_1 + k_2)} \left < \sum_{i = 1}^N u_i^{2k_1} \sum_{j = 1}^N u_j^{2k_2} \right >_\mathrm{conn} \nonumber \\
&= \frac{\Delta}{4} \log N + O(1)\,,\label{log-2pt}
\end{align}
independently of the value of $\nu$. Discarding terms that are not quartic in mass, this yields the free energy
\es{free-energy}{
F = -\frac{\Delta}{2} f_\text{log}^2 m^4 \log N + O(1) = -\frac{\Delta}{8} m^4 \log N + O(1)\,.
}
Note that for $\Delta=2$ this agrees with the result of \cite{Behan:2023fqq} for the $D_4$ theory, after setting $\mu_1=m$ and $\mu_2=\mu_3=\mu_4=0$. As we will see, the $O(1)$ terms are related to the $F^4$ vertex in the bulk effective action.\footnote{As seen in \eqref{f0-form}, the only positive powers of $u$ are independent of mass. This means there are no positive powers of $N$ in \eqref{free-energy} as expected from the form of the amplitude \eqref{Mintro}.} Computing them will require more information than what we have used so far.

We can get a flavor for how these future calculations might look by completing the expression for \eqref{universal-2pt}. The most important property of matrix models at large $N$ is that their eigenvalues become dense on a particular interval. One can therefore define the normalized density
\es{ev-density}{
\rho(x) = \frac{1}{N} \left < \sum_{i = 1}^N \delta(x - u_i) \right >, \quad \int_{-\nu}^\nu dx \, \rho(x) = 1\,.
}
To solve for $\nu$, consider the saddle point equation associated with \eqref{Zcanonical}
\es{saddle-eq}{
\int_{-\nu}^\nu dy \, \frac{\rho(y)}{x - y} = \frac{c^2_\mathcal{F}}{\Delta} x^{-1 + 2/\Delta}\,.
}
It has the solution
\es{saddle-sol}{
\rho(y) = \frac{c^2_\mathcal{F}}{\Delta} \frac{\sqrt{\nu^2 - y^2}}{\pi^2} \int_{-\nu}^\nu \frac{|x|^{-1 + 2/\Delta} \text{sgn}(x) - |y|^{-1 + 2/\Delta} \text{sgn}(y)}{x - y} \frac{dx}{\sqrt{\nu^2 - x^2}}\,,
}
which agrees with \cite{pastur99} after accounting for the different Vandermonde determinant. With the use of hypergeometric identities and some of the manipulations in \cite{av02}, \eqref{saddle-sol} evaluates to
\es{final-rho}{
\rho(y) = \frac{c^2_\mathcal{F} \sqrt{\nu^2 - y^2}}{\pi \nu^2 \Delta} {}_2F_1(1, 1 - 1/\Delta; 3/2; 1 - \nu^2/y^2)\,,
}
which reassuringly becomes the Wigner semicircle distribution for $\Delta \to 1$. The normalization condition then fixes
\es{final-nu}{
\nu = \left [ \frac{\sqrt{\pi} \Gamma(\Delta^{-1})}{c^2_\mathcal{F} \Gamma(\Delta^{-1} + 1/2)} \right ]^{\Delta / 2}\,,
}
which we introduced in \eqref{universal-2pt}.

\section{Consequences for F-theory}\label{sec:mellin}

Let us now turn to studying the structure of the moment map four-point function \eqref{momentmapfourpoint} in a large $N$ expansion in the various theories. In this regime, the correlator gives access to the study of gluon scattering on the singular locus AdS$_5\times S^3$ of the geometry \eqref{10dsol}, as first described in \cite{Alday:2021odx}. In this section we consider the expansion of this correlator at large $N$ in Mellin space for the theories listed in \eqref{sing}: after reviewing what is already known, we use the integrated correlator constraint introduced in Section \ref{sec:theories} together with the information derived in Section \ref{sec:localization} on the structure of the sphere free energy at large $N$ to fix a coefficient appearing at order $\log N/N^2$. This is related to the logarithmic terms appearing both in the gluon loop amplitude and in the graviton exchange amplitude, and we show that  the integrated constraint \eqref{integrated_constraint} fixes this term precisely in such a way as to match the prediction from the flat space limit.

\subsection{Gluon amplitudes in Mellin space}

Let us begin by reviewing what is known for the observable we are interested in. First of all, since results are most conveniently expressed in Mellin space we begin by introducing the Mellin representation of the reduced correlator as
\es{mellin_rep}{
\mathcal{G}_{\mathrm{conn}}^{A B C D}(U, V)=\int \frac{d s d t}{(4 \pi i)^2} U^{s / 2} V^{t / 2-2} M^{A B C D}(s, t) \Gamma[2-s / 2]^2 \Gamma[2-t / 2]^2 \Gamma[2-u / 2]^2\,,
}
where $s+t+u=6$ and $M^{A B C D}(s, t)$ is the Mellin amplitude, which can be expanded over projectors on the irreps in $\mathfrak{g}\otimes \mathfrak{g}$ as in \eqref{Girreps}, and the two integration contours are understood to include all poles in $s$ and $t$ but not those in $u$. In the limit of large $s$ and $t$, the Mellin amplitude $M(s,t)$ is expected to reproduce the scattering amplitude $\mathcal{A}(s,t)$ in the 8d flat space limit of the AdS$_5\times S^3$ supergravity background \cite{Penedones:2010ue}. More precisely,
\es{flatspace_def}{
\mathcal{A}^{A B C D}(s, t)=\lim _{L \rightarrow \infty} 8 \pi^4 L^4 \int \frac{d \beta}{2 \pi i} \frac{e^\beta}{\beta^4} \frac{L^4(u+s / w)^2}{16} M^{A B C D}\left(\frac{L^2}{2 \beta} s, \frac{L^2}{2 \beta} t\right)\,,
}
where the factor $(u+s/w)^2$ relates the Mellin transform of the reduced correlator $\mathcal{G}$ to that of the full correlator $G$ -- see \cite{Behan:2023fqq} for additional details.

In the large $N$ limit, we expect $M(s,t)$ to have an expansion of the form
\begin{align}\label{mellin_general}
M^{ABCD}(s,t)=&\frac{1}{N}M_{F^2}^{ABCD}(s,t)+\frac{1}{N^2}\left[M_R^{ABCD}(s,t)+M_{F^2|F^2}^{ABCD}(s,t)+b_{F^4} M_{F^4}^{ABCD}(s,t)\right] \nonumber\\
&+\frac{\log N}{N^2}\,b_{\log}\,M_{\log}^{ABCD}(s,t)+O(N^{-3}).
\end{align}
Let us now analyze each term independently. The tree-level gluon exchange term $M_{F^2}$ was computed in \cite{Alday:2021odx} and is given by
\es{gluontree}{
M_{F^2}^{ABCD}(s,t)=-\frac{4}{\Delta}\left[\frac{\mathtt{c}_s^{ABCD}}{(s-2)(u-2)}-\frac{\mathtt{c}_t^{ABCD}}{(t-2)(u-2)}\right]\,,
}
with its overall coefficient fixed in terms of the flavor central charge $k$ by the conformal Ward identities \cite{Osborn:1993cr}. The flavor structure in \eqref{gluontree} is completely fixed in terms of the tensors
\es{c_tree}{
\mathtt{c}_s^{ABCD}=f^{ABJ}f^{JCD}\,,\quad 
\mathtt{c}_t^{ABCD}=f^{ADJ}f^{JBC}\,,\quad 
\mathtt{c}_u^{ABCD}=f^{ACJ}f^{JDB}\,,
}
where for completeness we have also introduced $\mathtt{c}_u$ which is such that $\mathtt{c}_s+\mathtt{c}_t+\mathtt{c}_u=0$, due to the Jacobi identity. The one-loop gluon amplitude $M_{F^2|F^2}$ was computed in \cite{Alday:2021ajh} using the AdS unitarity method of \cite{Aharony:2016dwx}, which fixes the overall scale in terms of that of the tree-level amplitude \eqref{gluontree}. The result reads
\es{gluonloop}{
M_{F^2|F^2}^{ABCD}(s,t)=-\frac{1}{2\Delta^2}\left[\mathtt{d}_{st}^{ABCD}\mathcal{B}(s,t)+\mathtt{d}_{su}^{ABCD}\mathcal{B}(s,u)+\mathtt{d}_{tu}^{ABCD}\mathcal{B}(t,u)\right]\,,
}
where $s,t$ dependence of the function $\mathcal{B}(s,t)$ is given in \cite{Alday:2021ajh} and here it is normalized in such a way that for large $s,t$ it gives
\es{f_box}{
\mathcal{B}(s, t) & \simeq f_{\mathrm{box}}(s, t), \quad(s, t, \rightarrow \infty)\,, \\
f_{\mathrm{box}}(s, t) & =\frac{s t \log ^2\left(\frac{-s}{-t}\right)}{(s+t)^2}+2 \frac{s \log (-s)+t \log (-t)}{s+t}+\pi^2 \frac{s t}{(s+t)^2}\,,
}
while the flavor structures are given by
\es{d_loop}{
\mathtt{d}_{st}^{ABCD}&=f^{IAJ}f^{JBK}f^{KCL}f^{LDI}\,,\\
\mathtt{d}_{su}^{ABCD}&=f^{IAJ}f^{JBK}f^{KDL}f^{LCI}\,,\\
\mathtt{d}_{tu}^{ABCD}&=f^{IAJ}f^{JCK}f^{KBL}f^{LDI}\,.
}
The term $M_R$ represents the graviton exchange amplitude. As observed in \cite{Chester:2023qwo}, it contains the exchange of an infinite tower of Kaluza-Klein (KK) modes of the 10d graviton. Indeed, the decomposition of the one-half BPS multiplets of 4d $\cN=4$ superconformal symmetry to 4d $\cN=2$ will generate the 4d $\cN=2$ stress tensor, whose exchange amplitude was computed in \cite{Alday:2021ajh}, as well as an infinite tower of long multiplets. Computing the exchange amplitude for the latter would require the knowledge of their OPE coefficients, a problem which has not yet been tackled, so the full expression of $M_R$ is currently unavailable. The $M_{F^4}$ term represents the first higher-derivative correction, whose overall coefficient $b_{F^4}$ is unknown. The other unknown coefficient appearing in \eqref{mellin_general} is $b_{\log}$, which multiplies another 2-derivative contact term amplitude $M_{\log}$, appearing with a $\log N$ coefficient since its role is to regularize the logarithmic divergence of the one loop amplitude. The explicit expressions of $M_{F^4}$ and $M_{\log}$ are identical and read
\es{contact}{
M_{F^4}^{ABCD}(s,t)=M_{\log}^{ABCD}(s,t)=\delta^{AB}\delta^{CD}+\delta^{AC}\delta^{BD}+\delta^{AD}\delta^{BC}\,.
}
Note that the coefficients $b_{\log}$ and $b_{F^4}$ have been computed in \cite{Behan:2023fqq} for the $D_4$ theory, where the latter is a non-trivial function of the complexified gauge coupling $\tau$, while the former is a constant in the $D_4$ case as well, whose expression reads $b_{\log}=3$.\footnote{Note that because the full expression of $M_R$ is not known, it was only possible to compute $b_{F^4}(\tau)$ up to an additive constant, independent of $\tau$. Moreover, the structure of the contact terms in the $D_4$ theory is more complicated: the expressions given here only apply to $D_4$ after identifying triplets of representations permuted by triality, as explained in Appendix \ref{app:group}.}

In the rest of this section we will show how the sphere free energy can be used to constrain the value of the coefficient $b_{\log}$, while we leave the computation of $b_{F^4}$ for future work. This is due both to the fact that the full expression of the graviton exchange amplitude $M_R$ is currently unavailable and to the necessity of improving our understanding of the general matrix models presented in Section \ref{sec:localization}, as we explained there. We will then compare the flat space limit of the Mellin amplitude \eqref{mellin_general} with the expression of the flat space scattering amplitude, where the graviton exchange amplitude is in fact known, and show that our prediction for $b_{\log}$ matches the expectation from the flat space limit.

\subsection{Constraints from the sphere free energy}

Let us now consider the implications of the integrated correlator constraint \eqref{integrated_constraint} on the Mellin amplitude \eqref{mellin_general}. As explained, we shall focus solely on the $\log N/N^2$ term in \eqref{mellin_general}. The relevant terms in the expression of the mass deformed sphere free energy have been computed in \eqref{free-energy}, which allows us to obtain directly
\es{constraint_lhs}{
\left. -\partial^4_m F(m)\right|_{m=0}=3\Delta\,\log N+O(1)\,,
}
where the terms that we are not showing are $O(1)$ in the large $N$ expansion. While we do not know them for general $\Delta$, they were computed in \cite{Behan:2023fqq} for the $D_4$ theory, with $\Delta=2$. On the other hand, focusing on the $\log N/N^2$ term in \eqref{mellin_general} and specializing the flavor indices as dictated by \eqref{integrated_constraint}, we find\footnote{Note that specializing the flavor indices to $A=B=C=D=1$ simply corresponds to taking $M_{\log}=3$ in \eqref{contact}.}
\es{M1111log}{
\left. M^{1111}\right|_{\log N/N^2}=3\,b_{\log}\,.
}
The integral in \eqref{integrated_constraint} can then be conveniently performed using \cite{future}
\es{}{
I\left[M\right] \equiv-\int \frac{d s d t}{(4 \pi i)^2} & M(s, t) \Gamma[2-s / 2] \Gamma[s / 2] \Gamma[2-t / 2] \Gamma[t / 2] \Gamma[2-u / 2] \Gamma[u / 2] \\
& \times\left(\frac{H_{\frac{s}{2}-1}+H_{1-\frac{s}{2}}}{(t-2)(u-2)}+\frac{H_{\frac{t}{2}-1}+H_{1-\frac{t}{2}}}{(s-2)(u-2)}+\frac{H_{\frac{u}{2}-1}+H_{1-\frac{u}{2}}}{(s-2)(t-2)}\right)\,,
}
where $H_n$ is the $n-$th harmonic number, which is equivalent to \eqref{integral} once the Mellin representation \eqref{mellin_rep} is used. We find
\es{}{
I[1]=\frac{1}{24}\,,
}
which we can use with \eqref{M1111log} to compute the logarithmic term on the right hand side of \eqref{integrated_constraint}. Comparing the result of that integral with \eqref{constraint_lhs}, we obtain
\es{blog_fixed}{
b_{\log}=\frac{6}{\Delta}\,.
}
This is the main prediction of this paper. Note that in the $D_4$ theory, with $\Delta=2$, this matches the prediction $b_{\log}=3$ of \cite{Behan:2023fqq}.

\subsection{Constraints from the flat space limit}

Let us now consider the flat space limit of the Mellin amplitude \eqref{mellin_general} and compare it with the expectation from the known flat space amplitudes, which we discuss in Appendix \ref{app:flatspace}. The flat space action is given by the sum of the bulk supergravity action, where the dynamical fields are the ten-dimensional metric $G_{MN}$ and its superpartners, and the action for the 8d SYM theory living on the sevenbranes, which depends both on the eight-dimensional gluons $A_{\mu}$ and on the pull-back $g_{\mu\nu}$ of the ten-dimensional metric on the worldvolume of the sevenbranes (as well as the respective superpartners). The result can be expressed as
\es{S_total}{
S&=S_{\text{SUGRA}}[G_{MN},\ldots]+S_{\text{SYM}}[A_{\mu},\ldots;\,g_{\mu\nu},\ldots]\,,\\
S_{\text{SUGRA}}[G_{MN},\ldots]&=\frac{1}{2\kappa^2_{\mathrm{10d}}}\int d^{10}X\,\sqrt{-G}\,[R+\mathrm{HD}+\mathrm{SUSY}]\,,\\
S_{\text{SYM}}[A_{\mu},\ldots;\,g_{\mu\nu},\ldots]&=-\frac{1}{4\mathtt{g}^2_{\text{8d}}}\int d^8 x \sqrt{-g}\,\left[F^A_{\mu\nu}F^{A,\mu\nu}+\mathrm{HD}+\mathrm{SUSY}\right]\,,
}
where ``HD'' denotes higher-derivative terms in the low energy expansion of the effective action and ``SUSY'' denotes the supersymmetric completion of the given terms. In \eqref{S_total}, we have introduced the ten-dimensional gravitational coupling constant $\kappa_{\mathrm{10d}}$ and the eight-dimensional YM coupling constant $\mathtt{g}_{\text{8d}}$, which are related to the string parameters by\footnote{The result for $\mathtt{g}^2_{\text{8d}}$ can be derived starting from $\mathtt{g}^2_{\text{5d}}=8\pi^2/k$, which relates the gauge coupling of the AdS$_5$ gauge theory to the flavor central charge $k$ \cite{Freedman:1998tz}. Since the 8d and 5d gauge couplings are related by a dimensional reduction on the $S^3$ in \eqref{metricX5}, we then have $\mathtt{g}^{-2}_{\text{5d}}=L^4\text{vol}(S^3)\mathtt{g}^{-2}_{\text{8d}}$. Using $\text{vol}(S^3)=2\pi^2$, the result \eqref{centralcharges} for the flavor central charge $k$ and the expression \eqref{dictionary} for the AdS radius $L$, we finally obtain the expression for $\mathtt{g}^2_{\text{8d}}$ quoted in \eqref{kappagYM}.}
\es{kappagYM}{
2\kappa^2_{\mathrm{10d}}=(2\pi)^7g_s^2\ell_s^8\,,\quad
\mathtt{g}^2_{\text{8d}}=\frac{8\pi^4}{N\,\Delta}=(2\pi)^5g_s\ell_s^2\,.
}

From Appendix \ref{app:flatspace}, we learn that the structure of the flat space scattering amplitude is
\begin{align}\label{Aflat_general}
\mathcal{A}^{ABCD}(s,t)=&(u+s/w)^2\,\Big[g_s\ell_s^4\,\mathcal{A}_{F^2}^{ABCD}(s,t)+g_s^2\ell_s^8\left(\mathcal{A}_R+\mathcal{A}_{F^2|F^2}^{ABCD}(s,t)+\mathcal{A}_{F^4}^{ABCD}(s,t)\right)\nonumber\\
&\hspace{2.1cm}+g_s^2\ell_s^8\,\log \ell^2_s\,\mathcal{A}_{\log}^{ABCD}(s,t)+O(g_s^3\ell_s^{12})\Big]\,,
\end{align}
and now we discuss how this structure is reproduced by the flat space limit of the Mellin amplitude \eqref{mellin_general}, using the definition \eqref{flatspace_def} as well as the relations \eqref{kappagYM}. First, we find perfect agreement between the flat space limit of the gluon exchange amplitude \eqref{gluontree} and the corresponding flat space amplitude from Appendix \ref{app:flatspace}, both leading to
\es{Atree}{
\mathcal{A}_{F^2}^{ABCD}(s,t)=-\frac{(2\pi)^5}{stu}(t\,\mathtt{c}_s^{ABCD}-s\,\mathtt{c}_t^{ABCD})\,.
}
Similarly, the analytic dependence on $s,t$ of the gluon one-loop amplitude in flat space is reproduced by the flat space limit of $M_{F^2|F^2}$ in \eqref{gluonloop}, up to a renormalization constant which we do not determine here. The result is
\es{Aloop}{
\mathcal{A}_{F^2|F^2}^{ABCD}(s,t)=-\frac{2\pi^6}{3}\left[\mathtt{d}_{s t}^{ABCD} f_{\mathrm{box}}(s, t)+\mathtt{d}_{s u}^{ABCD} f_{\mathrm{box}}(s, u)+\mathtt{d}_{t u}^{ABCD} f_{\mathrm{box}}(t, u)\right]\,,
}
where $f_{\mathrm{box}}(s, t)$ is the function introduced in \eqref{f_box}. For the graviton exchange amplitude, while the AdS result $M_R$ is unkown as explained earlier in this section, we do have a flat space expectation which is computed in Appendix \ref{app:flatspace}. The fact that while gluons are constrained to live on the eight-dimensional worldvolume of the sevenbranes, gravitons can propagate in the whole ten-dimensional bulk, causes the presence of logarithmic terms in $\mathcal{A}_R$, which then inherits an analytic structure akin to that of a one loop amplitude, despite being a tree-level exchange \cite{Chester:2023qwo}. The result is
\es{AR}{
\mathcal{A}_R^{ABCD}(s,t)=-8\pi^6\,\Delta\,\left[
\delta^{AB}\delta^{CD}\,\log(-s)+
\delta^{AC}\delta^{BD}\,\log(-t)+
\delta^{AD}\delta^{BC}\,\log(-u)+\text{const.}\right]\,,
}
where the constant term is not relevant for our purposes. Finally, we do not consider the contact term $M_{F^4}$, for the reasons described earlier, while applying the flat space limit formula \eqref{flatspace_def} to $M_{\log}$ introduced in \eqref{contact} leads to
\es{Alog}{
\mathcal{A}_{\log}^{ABCD}(s,t)=-16\pi^6\,\Delta\,\left[
\delta^{AB}\delta^{CD}+
\delta^{AC}\delta^{BD}+
\delta^{AD}\delta^{BC}\right]\,,
}
where we have used the result \eqref{blog_fixed} to fix the coefficient $b_{\log}$ according to the prediction computed from the SW prepotential.

We are now ready to check that our prediction \eqref{blog_fixed} indeed agrees with the expectation in the flat space limit. This is possible because both the gluon loop amplitude $\mathcal{A}_{F^2|F^2}$ in \eqref{Aloop} and the graviton exchange amplitude $\mathcal{A}_R$ in \eqref{AR} contain logarithms of the Mandelstam variables: $\log(-s)$, etc. However, since the latter have dimension of energy squared, they are expected to combine with a suitable power of the string length $\ell_s$ to make the argument of such logarithmic terms dimensionless: $\log(-\ell^2_s\,s)$, etc. From the results \eqref{dfromt} and expressing all group-theoretic quantities of Appendix \ref{app:group} in terms of $\Delta$ using \eqref{delta_from_h}, it is straightforward to argue that 
\es{}{
\mathtt{d}_{st}^{ABCD}+\mathtt{d}_{su}^{ABCD}+\mathtt{d}_{tu}^{ABCD}=6\Delta\,\left[
\delta^{AB}\delta^{CD}+
\delta^{AC}\delta^{BD}+
\delta^{AD}\delta^{BC}\right]\,,
}
from which we find that the logarithmic terms in \eqref{Aloop}, \eqref{AR} and the coefficient in \eqref{Alog} are precisely such that the flat space amplitude \eqref{Aflat_general} can be rewritten as
\begin{align}\label{Aflat_new}
\mathcal{A}^{ABCD}(s,t)=&(u+s/w)^2\,\Big[g_s\ell_s^4\,\mathcal{A}_{F^2}+g_s^2\ell_s^8\left(\tilde{\mathcal{A}}_R^{ABCD}(s,t)+\tilde{\mathcal{A}}_{F^2|F^2}^{ABCD}(s,t)+\mathcal{A}_{F^4}^{ABCD}(s,t)\right) \nonumber\\
&\hspace{2.1cm}+O(g_s^3\ell_s^{12})\Big]\,,
\end{align}
where
\es{}{
\tilde{\mathcal{A}}_{F^2|F^2}^{ABCD}(s,t)&=-\frac{2\pi^6}{3}\left[\mathtt{d}_{s t}^{ABCD} \tilde{f}_{\mathrm{box}}(s, t)+\mathtt{d}_{s u}^{ABCD} \tilde{f}_{\mathrm{box}}(s, u)+\mathtt{d}_{t u}^{ABCD} \tilde{f}_{\mathrm{box}}(t, u)\right]\,,\\
\tilde{f}_{\mathrm{box}}(s, t) & =\frac{s t \log ^2\left(\frac{-s}{-t}\right)}{(s+t)^2}+2 \frac{s \log (-\ell_s^2\,s)+t \log (-\ell_s^2\,t)}{s+t}+\pi^2 \frac{s t}{(s+t)^2}\,,
}
and
\begin{align}
\tilde{\mathcal{A}}_R^{ABCD}(s,t)=&-8\pi^6\,\Delta\,\big[
\delta^{AB}\delta^{CD}\,\log(-\ell_s^2\,s)+
\delta^{AC}\delta^{BD}\,\log(-\ell_s^2\,t)+
\delta^{AD}\delta^{BC}\,\log(-\ell_s^2\,u)\big]\nonumber \\
&+\text{const.}\,,
\end{align}
where as anticipated the arguments of all logarithms are dimensionless. This provides a non-trivial check of the validity of our prediction \eqref{blog_fixed}.

\section{Conclusion}\label{sec:conclusion}

In this paper, we showed how the large $N$ expansion of the mass deformed sphere free energy $F(m)$ for certain non-Lagrangian theories with F-theoretic holographic duals can be written in terms of matrix model integrals using their Seiberg-Witten curves. These matrix models have a potential with fractional powers of the matrix eigenvalues, which differs from the matrix models that appear in standard localization calculations for Lagrangian theories. For each theory, we then computed the $\log N$ term from the matrix model and used it to fix the logarithmic threshold term in the holographic correlator of gluons on $AdS_5\times S^3$. In the flat space limit, our result matched the corresponding threshold term in the effective SYM theory of D7 branes in F-theory.

Our calculation is a first step toward using this matrix model description of $F(m)$ to further study stringy corrections to the low energy effective description of F-theory. The next step would be to study the $F^4$ correction to the effective SYM on the D7 branes by computing the leading large $N$ term in $F(m)$ from the matrix model, as was done using the matrix model that appears from supersymmetric localization for the Lagrangian $D_4$ theory in previous work \cite{Behan:2023fqq}. As a test case, we show in Appendix \ref{app_usp} that the non-instanton contributions to this term can also be computed from the Seiberg-Witten curve for the $D_4$ theory, and match the expected result from \cite{Behan:2023fqq}. The instanton terms may also be computable, but would require an infinite resummation. This calculation requires a certain analytic continuation of the naively divergent matrix model, which could be useful for the matrix models that appear in the other F-theoretic CFTs. For these other theories, we would require a more systematic understanding of large $N$ correlators in non-polynomial matrix models.

To fix the $F^4$ term, we not only need the leading large $N$ term in $F(m)$ but would also need to know the explicit form of the graviton exchange term $M_{R}$ that appears at the same order in $1/N$ in the holographic correlator.\footnote{The one-loop gluon term $M_{F^2|F^2}$ also appears at this order, but this was already computed in \cite{Alday:2021ajh}.} As discussed first in a related context in \cite{Chester:2023qwo}, the graviton exchange term is a sum over infinitely many Witten diagrams for the exchange of each single trace operator that comes from decomposing the graviton in 10d over $AdS_5\times S^3$. The relative coefficients of each diagram cannot be fixed from the conventional analytic bootstrap constraints, and require an explicit dimensional reduction of 10d supergravity to 8d SYM. If we can fix $F^4$ in AdS, then we could try to compare it to predictions in the flat space limit using duality to the heterotic string \cite{Lerche:1998gz}.\footnote{More precisely, the heterotic string is dual to Type IIB where the transverse space is a torus, while in the flat space limit of our holographic theory, the transverse space is flat. We would thus need to modify the results of \cite{Lerche:1998gz}, which should hopefully be possible since $F^4$ is protected.} 

More ambitiously, we can hope to study these F-theoretic CFTs also at finite $N$. For the $D_4$ theory, the theory with lowest $N$ is the same as superconformal SQCD with gauge group $SU(2)$ and complex coupling $\tau$, and was studied in \cite{Chester:2022sqb} as a function of $\tau$ by combining the numerical bootstrap with integrated constraints from $F(m)$ as computed using localization. Some of the other F-theoretic CFTs, namely those with flavor symmetry $SU(2)$ and $E_6$, were also studied using the numerical bootstrap in \cite{Cornagliotto:2017snu,Lemos:2015awa,Beem:2014zpa, Bissi:2021rei}, but without imposing any additional integrated constraints from $F(m)$.\footnote{Recall that \cite{Bissi:2021rei} compared a certain OPE coefficient computed from $F(m)$ to numerical bootstrap bounds, but did not impose it.} As discussed above, it is difficult to impose localization constraints via $F(m)$ for the other non-Lagrangian theories at finite $N$, since the Seiberg-Witten approach used in this paper seems to diverge at finite $N$ with no obvious regularization scheme. On the other hand, if we can understand how to compute $F(m)$ to all orders in $1/N$, then this could be used along with the numerical bootstrap to constrain the CFTs at large but finite $N$,\footnote{Note that constraints from the numerical bootstrap are always stronger than those from the analytic bootstrap even at large but finite values of $N$, because only the former non-trivially imposes unitarity.} where the large $N$ expansion may be a good approximation. If we can use these non-perturbative methods to study the large $N$ expansion of the holographic correlators, then this can be used to study stringy corrections to the F-theory effective action to any order.

\section*{Acknowledgments} 

We thank Xi Yin, Silviu Pufu, Bertrand Eynard, Ying-Hsuan Lin, Leonardo Rastelli, Christopher Beem, Cyril Closset, Davide Gaiotto, and Joseph McGovern for useful conversations. CB is supported by the S\~{a}o Paulo Research Foundation (FAPESP) grants 2023/03825-2 and 2019/24277-8. CB thanks the Perimeter Institute for Theoretical Physics for hospitality during part of this work. Research at Perimeter Institute is supported by the Government of Canada through Industry Canada and by the Province of Ontario through the Ministry of Research \& Innovation. SMC is supported by the Royal Society under the grant URF\textbackslash R1\textbackslash 221310. PF thanks the International Center for Theoretical Physics - South American Institute for Fundamental Research (ICTP-SAIFR) for hospitality during part of this work.

\appendix

\section{Group theory}\label{app:group}

It was first noticed by Deligne \cite{Deligne_1996} that the sequence of Lie algebras in Table \ref{tab:deligneseries}
\begin{table}[h!]
\centering
\begin{tabular}{|c || c | c | c | c | c|c|c|c|} 
\hline
$\mathfrak{g}$ & $A_1$ & $A_2$ & $G_2$ & $D_4$ & $F_4$ & $E_6$ & $E_7$ & $E_8$ \\ 
\hline
$h^{\vee}$ & 2 & 3 & 4 & 6 & 9 & 12 & 18 & 30 \\
\hline
\end{tabular}
\caption{Lie algebras in the Deligne series and their dual Coxeter numbers $h^{\vee}$.}
\label{tab:deligneseries}
\end{table}
enjoys certain uniform properties, namely the dimensions of the irreducible representations appearing in the $k$-fold tensor product of the adjoint representation can be expressed as a unique rational function (for each irrep) of the dual Coxeter number $h^\vee$. More evidence for this was later presented in \cite{CohenMan_1996}, whose notation we follow here. Note that the Lie algebras in Table \ref{tab:deligneseries} are precisely those corresponding to the flavor groups of the SCFTs relevant for this paper -- see \eqref{sing} -- except for $G_2$ and $F_4$.\footnote{The existence of 4d $\mathcal{N}=2$ SCFTs with flavor group $G_2$ and $F_4$ has been conjectured in \cite{Beem:2013sza} based on chiral algebra arguments. The issue is not yet completely settled.} The dimension $\text{dim}\,\mathfrak{g}$ of the Lie algebras in \ref{tab:deligneseries} can be expressed as a function of their dual Coxeter number $h^\vee$ as\footnote{We also note that restricting our attention to the algebras in \eqref{sing} (thus ignoring $G_2$ and $F_4$), we are only dealing with simply-laced Lie algebras, for which the Coxeter and dual Coxeter number agree. In those cases, the corresponding rank can also be expressed as a simple function of the dual Coxeter number as
\es{}{
\text{rank}\,\mathfrak{g}=\frac{\text{dim}\mathfrak{g}}{\h+1}=2\frac{5\h-6}{\h+6}\,.
}
}
\es{}{
\text{dim}\,\mathfrak{g}=2\frac{(\h+1)(5\h-6)}{\h+6}\,.
}
We are interested in the vector space $\mathfrak{g}\otimes \mathfrak{g}$, {\it i.e.} the tensor product of the adjoint representation with itself, which is relevant for four-point functions. This is a reducible representation of $\mathfrak{g}$, which can be written as a direct sum of irreducible $\mathfrak{g}$-modules as
\es{g2irreps}{
\mathfrak{g}\otimes \mathfrak{g}&=\text{Sym}^2\mathfrak{g}\oplus \wedge^2\mathfrak{g}\,,\\
\text{Sym}^2\mathfrak{g}&=\mathbf{1}\oplus \mathbf{Y_2^*}\oplus \mathbf{Y_2}\,,\\
\wedge^2\mathfrak{g}&= \mathfrak{g}\oplus \mathbf{X_2}\,,
}
where we are following the notation of \cite{CohenMan_1996} for the irreps, and $\text{Sym}^2$ denotes the symmetric part of $\mathfrak{g}\otimes \mathfrak{g}$ while $\wedge^2$ identifies its antisymmetric part. Besides the singlet and the adjoint, the dimensions of these representations, as a function of $h^\vee$, are \cite{CohenMan_1996}
\es{dimrepsXY}{
\text{dim}\,\mathbf{Y_2}&=\frac{5(\h)^2(2\h+3)(5\h-6)}{(\h+6)(\h+12)}\,,\\
\text{dim}\,\mathbf{Y_2^*} &=\frac{270(\h)^2(\h+1)(\h-2)}{(\h+6)^2(\h+12)}=\left. \text{dim}\,\mathbf{Y_2} \right|_{\h\to -\frac{6\h}{\h+6}}\,,\\
\text{dim}\,\mathbf{X_2}&=\frac{5(\h+1)(\h-2)(2\h+3)(5\h-6)}{(6+\h)^2}\,,
}
and for the interesting values of $\h$ the two sets of irreps in $\text{Sym}^2\mathfrak{g}$ and $\wedge^2\mathfrak{g}$ are listed in increasing order of dimension. We can also introduce a parity operator $(-1)^{|\mathbf{R}|}$ such that 
\es{}{
(-1)^{|\mathbf{R}|}=
\begin{cases}
+1\,,\quad \text{if}\quad \mathbf{R}\in \text{Sym}^2\mathfrak{g}\,,\\
-1\,,\quad \text{if}\quad \mathbf{R}\in \wedge^2\mathfrak{g}\,.
\end{cases}
}
A comment is in order for the two special cases of $A_1$ and $D_4$. We note that the adjoint representation of $A_1$ is the $\mathbf{3}$ and $\mathbf{3}\otimes \mathbf{3}=\mathbf{1}\oplus\mathbf{5}\oplus\mathbf{3}$, where the first two are symmetric and the last is antisymmetric. It is easy to check that for $\h=2$ we obtain  $\text{dim}\,\mathbf{Y_2^*}=\text{dim}\,\mathbf{X_2}=0$, so this explains why in that case there are only three representations. The case $D_4$ is a bit more subtle: the adjoint is $\mathbf{28}$ and
\es{}{
\mathbf{28}\otimes \mathbf{28}=\mathbf{1}\oplus \mathbf{35_v}\oplus \mathbf{35_c}\oplus \mathbf{35_s}\oplus \mathbf{300}\oplus \mathbf{28}\oplus \mathbf{350}\,,
}
where there are three $\mathbf{35}$ representations permuted by triality. We can still make sense of \eqref{g2irreps} in this case simply by identifying $\mathbf{Y_2^*}$ with the reducible representation $\mathbf{105}=\mathbf{35_v}\oplus \mathbf{35_c}\oplus \mathbf{35_s}$ (note that indeed $\text{dim}\mathbf{Y_2^*}=105$ for $\h=6$). 

Keeping these two caveats in mind, we can proceed by identifying a convenient basis of five tensors that span the vector space $\mathfrak{g}\otimes \mathfrak{g}$, in terms of which we can express the various tensor structures that enter in the Mellin amplitudes, as well as the projectors on the various irreps. We find that a convenient basis is
\es{basistensors}{
\mathtt{t}_1^{ABCD}&=\delta^{AB}\delta^{CD}\,,\quad
\mathtt{t}_2^{ABCD}=\delta^{AC}\delta^{DB}\,,\quad
\mathtt{t}_3^{ABCD}=\delta^{AD}\delta^{BC}\,,\\
\mathtt{t}_4^{ABCD}&=f^{ADJ}f^{JBC}\,,\quad
\mathtt{t}_5^{ABCD}=f^{ACJ}f^{JDB}\,,
}
where we have introduced the structure constants $f^{ABC}$ via
\es{}{
[T^A,T^B]=i\,f^{ABC}\,T^C\,,
}
and we normalize our generators in such a way that the square norm $\psi^2$ of the longest root of each Lie algebra is 2:\footnote{Note that all Lie algebras in \eqref{sing} are simply laced, so in those cases their roots all have the same length.}
\es{}{
\psi^2=2\,.
}
It is useful to express the flavor structures appearing in the Mellin amplitudes of Section \ref{sec:mellin} in terms of the basis tensors \eqref{basistensors}. For the tree-level structures defined in \eqref{c_tree} we find trivially,
\es{cfromt}{
\mathtt{c}_s=-\mathtt{t}_4-\mathtt{t}_5\,,\quad
\mathtt{c}_t=\mathtt{t}_4\,,\quad
\mathtt{c}_u=\mathtt{t}_5\,.
}
The expressions for the one-loop structures \eqref{d_loop} are slightly more complicated to derive for the case of exceptional groups, but using the results of \cite{Marrani:2010de} (see in particular eq. (2.18)) we are able to find, for all Lie algebras in Table \ref{tab:deligneseries},
\es{dfromt}{
\mathtt{d}_{st}&=\frac{\psi^2h^{\vee}}{6}\left[\frac{5\psi^2\,h^{\vee}}{2+\text{dim}\,\mathfrak{g}}(\mathtt{t}_1+\mathtt{t}_2+\mathtt{t}_3)-(\mathtt{c}_s-\mathtt{c}_t)\right]\,,\\
\mathtt{d}_{su}&=\frac{\psi^2h^{\vee}}{6}\left[\frac{5\psi^2\,h^{\vee}}{2+\text{dim}\,\mathfrak{g}}(\mathtt{t}_1+\mathtt{t}_2+\mathtt{t}_3)-(\mathtt{c}_u-\mathtt{c}_s)\right]\,,\\
\mathtt{d}_{tu}&=\frac{\psi^2h^{\vee}}{6}\left[\frac{5\psi^2\,h^{\vee}}{2+\text{dim}\,\mathfrak{g}}(\mathtt{t}_1+\mathtt{t}_2+\mathtt{t}_3)-(\mathtt{c}_t-\mathtt{c}_u)\right]\,,
}
which we have written in terms of $\mathtt{c}_s$, $\mathtt{c}_t$ and $\mathtt{c}_u$ to emphasize the permutation symmetry but they can of course be written in terms of the basis tensors \eqref{basistensors} using \eqref{cfromt}.

The next step is to identify the projectors $P_{\mathbf{R}}^{ABCD}$ on each irrep $\mathbf{R}$ appearing in \eqref{g2irreps}, which satisfy the properties
\es{ProjectorsProperties}{
P^{ABCD}_{\mathbf{R}}&=P^{CDAB}_{\mathbf{R}}\,,\\
P^{ABCD}_{\mathbf{R}}&=(-1)^{|\mathbf{R}|}\,P^{BACD}_{\mathbf{R}}\,,\\
P^{ABEF}_{\mathbf{R}_1}P^{FECD}_{\mathbf{R}_2}&=\delta_{\mathbf{R}_1,\mathbf{R}_2}\,P^{ABCD}_{\mathbf{R}_1}\,,\\
P^{ABCD}_{\mathbf{R}_1}P^{ABCD}_{\mathbf{R}_2}&=\delta_{\mathbf{R}_1,\mathbf{R}_2}\,\text{dim}(\mathbf{R}_1)\,,
}
and are solutions to the Casimir equation
\es{}{
\mathtt{c}_s^{EABF}P^{EFCD}_{\mathbf{R}}=-\lambda({\mathbf{R}})\,P^{ABCD}_{\mathbf{R}}\,.
}
The projectors can be found by making an ansatz in terms of a linear combination of the tensors in \eqref{basistensors} and fixing the coefficients and the eigenvalue $\lambda({\mathbf{R}})$ with the Casimir equation and requiring the properties \eqref{ProjectorsProperties}. To obtain the results, it is useful to employ the obvious properties of \eqref{basistensors} under permutations of their indices, as well as
\es{}{
\mathtt{t}_a^{ABEF}\mathtt{t}_b^{EFCD}=
\begin{pmatrix}
\text{dim}\,\mathfrak{g}\,\mathtt{t}_1^{ABCD} &  \mathtt{t}_1^{ABCD} & \mathtt{t}_1^{ABCD} & \psi^2h^{\vee}\mathtt{t}_1^{ABCD} & -\psi^2h^{\vee}\mathtt{t}_1^{ABCD}\\
\mathtt{t}_1^{ABCD} & \mathtt{t}_2^{ABCD} & \mathtt{t}_3^{ABCD} & \mathtt{t}_4^{ABCD} & \mathtt{t}_5^{ABCD}\\
\mathtt{t}_1^{ABCD} & \mathtt{t}_3^{ABCD} & \mathtt{t}_2^{ABCD} & -\mathtt{t}_5^{ABCD} & -\mathtt{t}_4^{ABCD}\\
\psi^2h^{\vee}\mathtt{t}_1^{ABCD} & \mathtt{t}_4^{ABCD} & -\mathtt{t}_5^{ABCD} &\mathtt{d}_{su}^{ABCD} & -\mathtt{d}_{st}^{ABCD}\\
-\psi^2h^{\vee}\mathtt{t}_1^{ABCD} & \mathtt{t}_5^{ABCD} & -\mathtt{t}_4^{ABCD} &-\mathtt{d}_{st}^{ABCD}  & \mathtt{d}_{su}^{ABCD} 
\end{pmatrix}\,,
}
from which the action of the Casimir on $\mathtt{t}_a$ can be easily argued, as well as
\es{}{
\mathtt{t}_a^{ABCD}\mathtt{t}_b^{ABCD}=
\text{dim}\,\mathfrak{g}
\begin{pmatrix}
\text{dim}\,\mathfrak{g} & 1 & 1 &\psi^2h^{\vee} &-\psi^2h^{\vee}  \\
1 &\text{dim}\,\mathfrak{g} & 1 &-\psi^2h^{\vee} & 0 \\
1 & 1 & \text{dim}\,\mathfrak{g} &0 & \psi^2h^{\vee} \\
\psi^2h^{\vee} & -\psi^2h^{\vee} & 0 &(\psi^2h^{\vee})^2 & -\tfrac{1}{2}(\psi^2h^{\vee})^2 \\
-\psi^2h^{\vee} & 0 & \psi^2h^{\vee} &-\tfrac{1}{2}(\psi^2h^{\vee})^2 & (\psi^2h^{\vee})^2 \\
\end{pmatrix}\,,
}
which is needed to enforce the correct normalization of the projectors. We find that the eigenvalues corresponding to the representations appearing in \eqref{g2irreps} are
\es{}{
\lambda(\mathbf{1})=\psi^2\,\h\,,\quad
\lambda(\mathbf{Y_2^*})=\frac{\psi^2}{6}\,(\h+6)\,,\quad
\lambda(\mathbf{Y_2})=-\psi^2\,,\quad
\lambda(\mathfrak{g})=0\,,\quad
\lambda(\mathbf{X_2})=\frac{\psi^2\,\h}{2}\,.
}
We are now ready to give the expressions for the corresponding projectors in terms of the basis tensors $\mathtt{t}_a$, which read
\es{projectors}{
P_{\mathbf{1}}&=\frac{1}{\text{dim}\,\mathfrak{g}}\left(1,\,0,\,0,\,0,\,0\right)\cdot \mathtt{t}\,,\\
P_{\mathbf{Y_2^*}}&=\frac{1}{\h+12}\left(-\frac{6(\h+1)}{\text{dim}\,\mathfrak{g}},\,3,\,3,\,\frac{3}{2},\,-\frac{3}{2}\right)\cdot \mathtt{t}\,,\\
P_{\mathbf{Y_2}}&=\frac{1}{\h+12}\left(\frac{(5\h-6)}{\text{dim}\,\mathfrak{g}},\,\frac{\h+6}{2},\,\frac{\h+6}{2},\,-\frac{3}{2},\,\frac{3}{2}\right)\cdot \mathtt{t}\,,\\
P_{\mathfrak{g}}&=-\frac{1}{2\h}\left(0,0,0,-1,-1\right)\cdot \mathtt{t}\,,\\
P_{\mathbf{X_2}}&=+\frac{1}{2\h}\left(0,\h,-\h,1,1\right)\cdot \mathtt{t}\,.
}
We can also define flavor crossing matrices which relate the projectors with different permutations of indices as
\es{}{
(F_t)_a^{\,\,\,b}=\frac{1}{\text{dim}\,\mathbf{R}_a}P_{\mathbf{R}_a}^{CBAD}P_{\mathbf{R}_b}^{ABCD}\,,\quad
(F_u)_a^{\,\,\,b}=\frac{1}{\text{dim}\,\mathbf{R}_a}P_{\mathbf{R}_a}^{DBCA}P_{\mathbf{R}_b}^{ABCD}\,,
}
where note that
\es{}{
(-1)^{|\mathbf{R}_a|}(F_t)_a^{\,\,\,b}=(-1)^{|\mathbf{R}_b|}(F_t)_a^{\,\,\,b}\,.
}
We can therefore limit to giving one of the two matrices and we choose $F_u$, which reads
\es{}{
F_u=
\begin{pmatrix}
\frac{1}{\text{dim}\,\mathfrak{g}} & \frac{\text{dim}\,\mathbf{Y_2^*}}{\text{dim}\,\mathfrak{g}} & \frac{\text{dim}\,\mathbf{Y_2}}{\text{dim}\,\mathfrak{g}} & 1 & \frac{\text{dim}\,\mathbf{X_2}}{\text{dim}\,\mathfrak{g}} \\
\frac{1}{\text{dim}\,\mathfrak{g}} & -\frac{5(\h)^2+36}{2(\h+1)(5\h-6)} & \frac{(\h+6)(2\h+3)}{2(\h+1)(\h+12)} & \frac{\h+6}{6} & -\frac{2\h+3}{3\h} \\
\frac{1}{\text{dim}\,\mathfrak{g}} & \frac{27(\h-2)}{(\h+12)(5\h-6)} &  \frac{(\h)^2+2\h+6}{2(\h+12)(\h+1)}& - \frac{1}{\h} & - \frac{\h-2}{2\h} \\
\frac{1}{\text{dim}\,\mathfrak{g}} &\frac{45\h(\h-2)}{2(\h+12)(5\h-6)} & -\frac{5\h(2\h+3)}{2(\h+12)(\h+1)} & \frac{1}{2} & 0 \\
\frac{1}{\text{dim}\,\mathfrak{g}} & -\frac{18\h}{(\h+12)(5\h-6)} & -\frac{\h(\h+6)}{2(\h+12)(\h+1)} & 0 & \frac{1}{2} \\
\end{pmatrix}\,,
}
where the rows and columns run over the irreducible representations in $\mathfrak{g}\otimes \mathfrak{g}$, listed in the order of appearance in \eqref{projectors}. Note that, after using the various identities presented in this appendix, the expressions of the projectors in terms of the basis tensors \eqref{basistensors}, as well as that of the flavor crossing matrices, only depend on the dual Coxeter number $\h$. This is true for all Lie algebras in Table \ref{tab:deligneseries}.

Finally, in Section \ref{sec:theories} we have discussed the relation between the mass deformed sphere free energy and a certain integral of the component of the reduced correlator $\mathcal{G}^{ABCD}$ with identical indices $A=B=C=D=1$. We can now express $\mathcal{G}^{1111}$ in terms of the irreducible representations in \eqref{g2irreps} as
\es{}{
\mathcal{G}^{1111}=\frac{1}{\text{dim}\mathfrak{g}}\left[\mathcal{G}_{\mathbf{1}}+\frac{\h+6}{5(\h)^2}\left(\text{dim} \mathbf{Y_2}\,\mathcal{G}_{\mathbf{Y_2}}+\text{dim}\mathbf{Y_2^*}\,\mathcal{G}_{\mathbf{Y_2^*}}\right)\right]\,,
}
while in terms of the basis tensors \eqref{basistensors} $\mathcal{G}^{1111}$ is simply obtained setting $\mathtt{t}_1=\mathtt{t}_2=\mathtt{t}_3=1$ and $\mathtt{t}_4=\mathtt{t}_5=0$.

\section{Checks of the $D_4$ theory}\label{app_usp}

\subsection{Expanding around large eigenvalues}
In Section \ref{sec:localization}, we treated correlation functions like $\left < \text{tr} \, u^{2k_1} \dots \text{tr} \, u^{2k_n} \right >_\mathrm{conn}$ (to a given order in $1/N$) as analytic functions of the exponents. This is what led to the crucial result that the $\log N$ contribution to the free energy came from a single two-point function as in \eqref{log-2pt}. In this section, we will apply this idea to the $D_4$ theory which is $USp(2N)$ SQCD with four fundamental hypers and one antisymmetric hyper. This provides a non-trivial check since the perturbative terms in the sphere free energy for this theory were found independently in \cite{Beccaria:2021ism,Beccaria:2022kxy,Behan:2023fqq} by using the convergent weak coupling expansion.

Neglecting the contribution of the instantons, the matrix model has the expression\footnote{The eigenvalues $a_n$ here are real and they were called $x_n$ in \cite{Behan:2023fqq}. Other papers such as \cite{bfgl11} use the convention that the $a_n$ are imaginary. For the $D_4$ theory at zero mass, $u_n = a_n^2/8$.}
\es{ZUSp}{
Z(\mu_i) &= \int [dA] \, e^{-\frac{8\pi^2}{g^2_\text{YM}} \text{tr} A^2} \prod_{n = 1}^N \frac{H(2a_n)^2}{\prod_{j = 1}^4 H(a_n + \mu_j) H(a_n - \mu_j)} \\
&= \int [dA] \, e^{-\frac{8\pi^2 N}{\lambda} \text{tr} A^2 - S^{(0)}_\text{int} - m_{(2)} S^{(2)}_\text{int} - m_{(4)} S^{(4)}_\text{int} + O(\mu^6)}\,,
}
where $[dA]$ is given by \eqref{Zmeasure},
\es{ir-coupling}{
\frac{1}{\lambda} = \frac{1}{g^2_\text{YM} N} + \frac{\log 2}{2\pi^2 N}\,,
}
is the IR coupling and $H(a) \equiv e^{-(1 + \gamma)a^2} G(1 + ia)G(1 - ia)$. Of the three quartic mass invariants
\es{so8-mass}{
m_{(4)} \equiv \frac{1}{4} \sum_{i = 1}^4 \mu_i^4\,, \quad m_{(2)}^2 \equiv \left ( \frac{1}{4} \sum_{i = 1}^4 \mu_i^2 \right )^2\,, \quad m_\text{Pf} \equiv \mu_1 \mu_2 \mu_3 \mu_4\,,
}
associated with the three quartic Casimirs of $SO(8)$, the last one is only generated by instantons. Instead of considering small eigenvalues as in \cite{Beccaria:2021ism,Beccaria:2022kxy,Behan:2023fqq}, we will use a well known asymptotic expansion of the Barnes G function to write
\es{logH-asymp}{
\log H(a) = \frac{1}{6} \left [ 1 - 12 \log \mathtt{A} + 3(1 - 2\gamma)a^2 - (1 +6a^2) \log a \right ] - 2 \sum_{g = 2}^\infty \frac{(-1)^gB_{2g}}{2g(2g - 2) a^{2g - 2}}\,,
}
where $\mathtt{A}$ is the Glaisher–Kinkelin constant and $B_{n}$ is the $n-$th Bernoulli number. This allows us to obtain
\es{int-actions}{
S^{(0)}_\text{int} &= \sum_{n = 1}^N \left [ -\log a_n + \frac{1}{3} \left ( \log 2 + 3 - 36\log \mathtt{A} \right ) + \frac{1}{16 a_n^2} + O(a_n^{-4}) \right ]\,, \\
S^{(1)}_\text{int} &= \sum_{n = 1}^N \left [ -8\log a_n + \frac{2}{3 a_n^2} + O(a_n^{-4}) \right ]\,, \\
S^{(2)}_\text{int} &= \sum_{n = 1}^N \left [ \frac{2}{3 a_n^2} + O(a_n^{-4}) \right ]\,.
}
Computing the sphere free energy to order $1/N$ via the cumulant expansion will require connected correlation functions with up to three traces. The results for these are given in \cite{Beccaria:2021ism} as
\begin{align}
& \left < \mathrm{tr} \, a^{2k} \right > = N \left ( \frac{\lambda}{8\pi^2} \right )^k \frac{2^{k + 1} \Gamma[k + \tfrac{1}{2}]}{\sqrt{\pi} \Gamma[k + 2]} \left [ 1 + \frac{k + 1}{4N} + \frac{k(k^2 - 1)}{48N^2} \right ] + O(N^{-2})\,, \label{corr-results} \\
& \left < \mathrm{tr} \, a^{2k_1} \mathrm{tr} \, a^{2k_2} \right >_\mathrm{conn} = \left ( \frac{\lambda}{8\pi^2} \right )^{k_1 + k_2} \frac{2^{k_1 + k_2 + 1} \Gamma[k_1 + \tfrac{1}{2}] \Gamma[k_2 + \tfrac{1}{2}]}{\pi (k_1 + k_2) \Gamma[k_1] \Gamma[k_2]} \left [ 1 + \frac{k_1 + k_2}{4N} \right ] + O(N^{-2})\,, \nonumber \\
& \left < \mathrm{tr} \, a^{2k_1} \mathrm{tr} \, a^{2k_2} \mathrm{tr} \, a^{2k_3} \right >_\mathrm{conn} = \frac{1}{N} \left ( \frac{\lambda}{8\pi^2} \right )^{k_1 + k_2 + k_3} \frac{2^{k_1 + k_2 + k_3 + 1} \Gamma[k_1 + \tfrac{1}{2}] \Gamma[k_2 + \tfrac{1}{2}] \Gamma[k_3 + \tfrac{1}{2}]}{\pi^{3/2} \Gamma[k_1] \Gamma[k_2] \Gamma[k_3]} + O(N^{-2})\,. \nonumber
\end{align}
We have not gone beyond $O(a_n^{-2})$ in \eqref{int-actions} because all such terms give zero when the correlation functions in \eqref{corr-results} are analytically continued. This is because the higher-point functions (and the one-point function after order $1/N$) always have a $\Gamma[k_i]$ in the denominator and its divergence for $k_i \leq 0$ is only cancelled in special cases. More precisely, the following conditions dictate which terms need to be considered.
\begin{enumerate}
\item Order $N$ contributions vanish when $k \leq -2$.
\item Order $1$ contributions from the one-point function vanish when $k \leq -1$. Order $1$ contributions from the two-point function vanish when at least one exponent is non-positive and $k_1 + k_2 \neq 0$.
\item Order $1/N$ contributions vanish when at least one exponent is non-positive.
\end{enumerate}
The leading large $N$ part of the two-point function is therefore the only part of \eqref{corr-results} where the large eigenvalue expansion of one operator is not guaranteed to truncate. Arbitrarily negative powers of $a$ are allowed as long as they are paired with equal and opposite powers of $a$ from the other operator. These terms, which are always $O(N^0 \lambda^0)$, are the most difficult to account for because they will be missed if we try to take the large eigenvalue expansion of both operators.

With this caveat in mind, the only three correlators that can contribute are
\es{nonzero-corrs}{
\left < \text{tr} \, a^{-2} \right > &= -\frac{16 N \pi^2}{\lambda}\,, \\
\left < \text{tr} \, \log a \, \text{tr} \, \log a \right >_\mathrm{conn} &= \frac{1}{2} \log \lambda - \log \pi - 2 \log 2 + \frac{1}{8N}\,, \\
\left < \text{tr} \, \log a \, \text{tr} \, \log a \, \text{tr} \, \log a \right >_\mathrm{conn} &= \frac{1}{4N}\,,
}
which do not have any perturbative large $N$ corrections. We can therefore compute the $O(\mu^4)$ contribution to the perturbative free energy defined in \cite{Behan:2023fqq} as.
\begin{align}
F_\text{pert} =& m_{(4)} \left [ \left < S^{(2)}_\text{int} \right > - \left < S^{(2)}_\text{int} S^{(0)}_\text{int} \right > + \frac{1}{2} \left < S^{(2)}_\text{int} S^{(0)}_\text{int} S^{(0)}_\text{int} \right > \right ]  \nonumber\\
&- \frac{m_{(2)}^2}{2} \left [ \left < S^{(1)}_\text{int} S^{(1)}_\text{int} \right > - \left < S^{(0)}_\text{int} S^{(1)}_\text{int} S^{(1)}_\text{int} \right > \right ] + O(N^{-2})  \\
=& -m_{(4)} \frac{16\pi^2 N}{3\lambda} + 2 m_{(2)}^2 \left [ 2\log \left ( \frac{16\pi^2}{\lambda} \right ) - \frac{1}{N} \right ] + O(N^{-2})\nonumber\,. \label{fpert-large}
\end{align}
Comparing this to the previously known result,
\es{fpert-small}{
F_\text{pert} = -\frac{m_{(4)}}{3}  \left [ \frac{16\pi^2 N}{\lambda} - 8\log 2 \right ] + 2 m_{(2)}^2 \left [ 2\log \left ( \frac{16\pi^2}{\lambda} \right ) - \frac{22}{3} - 4\gamma + 4\zeta(3) - \frac{1}{N} \right ] + O(N^{-2})\,,
}
and the two indeed agree up to $O(N^0 \lambda^0)$ terms.

\subsection{Considering instanton terms}
The previous subsection has shown that the large eigenvalue expansion, which is essential for the $A_1$, $A_2$ and $E_n$ theories considered here, gives the same perturbative free energy as the small eigenvalue expansion for $D_4$. More ambitiously, one could attempt to fully reproduce the results of \cite{Behan:2023fqq} with the methods advocated in the present paper which make no reference to a Lagrangian description. The following steps would be involved.
\begin{enumerate}
\item Expanding the prepotential around flat space instead of instanton-by-instanton.
\item Finding the prepotential to be expanded from the Seiberg-Witten curve instead of the Nekrasov partition function.
\end{enumerate}
In this subsection, we will briefly outline the first of these two steps which has implications for the protected vertices not considered in this paper like $F^4$.

Restoring the radius, \eqref{ZUSp} turns into
\es{ZUSp_deformed}{
Z(\mu_i) = \int [dA] \, e^{-\frac{8\pi^2 r^2}{g^2_\text{YM}} \text{tr} A^2} \prod_{n = 1}^N \frac{H(2r a_n)^2}{\prod_{j = 1}^4 H(r a_n + r \mu_j) H(r a_n - r \mu_j)} | Z_\text{inst}(rA, r\mu_i, q) |^2.
}
which makes it clear what happens with zero instantons.\footnote{A novel term called $Z_\text{extra}$ was needed for the approach in \cite{Behan:2023fqq} but it can be neglected here. Without using any of its detailed properties, dimensional analysis ensures that it produces negative powers of $\epsilon_{1, 2}$ which therefore do not enter the prepotential to a fixed order in $r^{-2}$.} The large radius expansion simply becomes equivalent to the large eigenvalue expansion because of how $r$ and $a_n$ appear together. To check what happens with instantons, we can expand the Nekrasov partition function at large radius and upgrade the $q$ dependent coefficients of $a_n^{-2k}$ into quasi-modular functions of weight $2k$.\footnote{To ensure $SL(2, \mathbb{Z})$ invariance, these are either Eisenstein series or Jacobi theta functions depending on how the associated mass combinations transform under $SO(8)$ triality.} In terms of the masses from \cite{Seiberg:1994aj}
\es{so8-mass2}{
\mathtt{N} &= \frac{3}{16} \sum_{i < j < k} \mu_i^2 \mu_j^2 \mu_k^2 - \frac{1}{96} \sum_{i \neq j} \mu_i^2 \mu_j^4 - \frac{1}{96} \sum_{i = 1}^4 \mu_i^6, \\
\mathtt{T}_2 &= -\frac{1}{24} \sum_{i < j} \mu_i^2 \mu_j^2 + \frac{1}{48} \sum_{i = 1}^4 \mu_i^4 - \frac{1}{2} \mu_1 \mu_2 \mu_3 \mu_4, \\
\mathtt{T}_1 &= \frac{1}{12} \sum_{i < j} \mu_i^2 \mu_j^2 - \frac{1}{24} \sum_{i = 1}^4 \mu_i^4, \quad \mathtt{R} = \frac{1}{2} \sum_{i = 1}^4 \mu_i^2,
}
this leads to the prepotentials
\begin{align}
\mathcal{F}_0 &= \frac{1}{2} \log q \sum_n a_n^2 - 2\mathtt{R} \sum_n \log a_n + \left ( \mathtt{T}_1 \theta_4^4 - \mathtt{T}_2 \theta_2^4 - \frac{\mathtt{R} E_2}{6} \right ) \sum_n a_n^{-2}\nonumber \\
&+ \left [ \frac{\mathtt{R}^3}{180}(5E_2^2 + E_4) + \frac{\mathtt{N}}{5} E_4 - \frac{\mathtt{R} \mathtt{T}_1}{6} \theta_4^4 (2E_2 + 2\theta_2^4 + \theta_4^4) + \frac{\mathtt{R} \mathtt{T}_2}{6} \theta_2^4 (2E_2 - 2\theta_4^4 - \theta_2^4) \right ] \sum_n a_n^{-4} + O(a_n^{-6})\,, \nonumber \\
\mathcal{F}_1 &= \frac{3}{2} \sum_n \log a_n + \sum_{m < n} \log(a_m^2 - a_n^2) + \frac{\mathtt{R} E_2}{6} \sum_n a_n^{-2} \nonumber \\
&- \frac{1}{2} \left [ \frac{\mathtt{R}^2}{18} (E_2^2 + 2E_4) - \mathtt{T}_1 \theta_4^4 (2\theta_2^4 + \theta_4^4) - \mathtt{T}_2 \theta_2^4 (2\theta_4^4 + \theta_2^4) \right ] \sum_n a_n^{-4} \nonumber \\
&- \frac{1}{3} \left [ \frac{\mathtt{R}^2}{12} (E_4 - E_2^2) + \mathtt{T}_1 \theta_4^4 (E_2 - 2\theta_2^4 - \theta_4^4) - \mathtt{T}_2 \theta_2^4 (E_2 + 2\theta_4^4 + \theta_2^4) \right ] \sum_{m < n} a_m^{-2} a_n^{-2} + O(a_n^{-6})\,, \nonumber \\
\mathcal{F}_2 &= -\frac{E_2}{32} \sum_n a_n^{-2} + \frac{\mathtt{R}}{960} (5E_2^2 + 43E_4) \sum_n a_n^{-4} + \frac{\mathtt{R}}{36} (E_4 - E_2^2) \sum_{m < n} a_m^{-2} a_n^{-2} + O(a_n^{-6})\,, \nonumber \\
\mathcal{F}_3 &= -\frac{E_4}{128} \sum_n a_n^{-4} + \frac{1}{192} (E_2^2 - E_4) \sum_{m < n} a_m^{-2} a_n^{-2} + O(a_n^{-6}). \label{many-prepot}
\end{align}
The rank one versions agree with the results of \cite{bfgl11, Billo:2013fi}. Further note that $\mathcal{F}_0$ indeed starts with the term $-\frac{1}{2} m^2 \log u$ when only one mass is turned on.\footnote{A further check is that these prepotentials obey the anomaly equation $\partial_{E_2} \mathcal{F}_g = \frac{1}{24} (\partial_a^2 \mathcal{F}_{g - 1} + \sum_{h = 1}^{g - 1} \partial_a \mathcal{F}_h \cdot \partial_a \mathcal{F}_{g - h})$ from \cite{Bershadsky:1993ta,Huang:2006si,Huang:2009md,Krefl:2010fm,hkk11}. This has previously been checked for $\mathcal{N} = 4$ SYM with any gauge group and $SU(2)$ SQCD \cite{Billo:2013fi, Billo:2013jba, bfflmpp14, Billo:2015pjb, Billo:2015jyt}. See also \cite{Ashok:2015cba} for theories with a duality group other than $SL(2, \mathbb{Z})$.} The key property of \eqref{many-prepot} is the repeated appearance of multi-trace terms with the expectation values
\es{problematic-term}{
\frac{1}{N^{g + 1}} \left < \sum_{i_1 < \dots < i_{g + 1}} a_{i_1}^{-2} \dots a_{i_{g + 1}}^{-2} \right > \approx \frac{1}{(g + 1)! N^{g + 1}} \left < \sum_{i = 1}^N a_i^{-2} \right >^{g + 1} = O(1)\,.
}
This suggests that all $\mathcal{F}_g$ are equally important for capturing instanton effects. Finding a way around this to study protected vertices will be an important problem for the future.

\section{Flat space amplitudes}\label{app:flatspace}

In this appendix we compute the flat space scattering amplitudes associated with the various terms in the expansion \eqref{Aflat_general}, which we have used in Section \ref{sec:mellin} to verify our prediction for the logarithmic threshold $b_{\log}$ in the Mellin amplitude \eqref{mellin_general}. The derivation of the amplitudes follows the same logic as Appendix B of \cite{Behan:2023fqq}, so we just review the main results here. The tree-level and one-loop pure gluon amplitudes can be argued from the results of \cite{Bern:1998ug} and they read
\es{A4tree}{
\mathcal{A}^{ABCD}_{4,\mathrm{gluon-exch}}(s,t)=-\mathtt{g}^2_{\text{8d}}\,\frac{(u+s/w)^2}{stu}\left(t\,\mathtt{c}_s^{ABCD}-s\,\mathtt{c}_t^{ABCD}\right)\,
}
and 
\es{A4loop}{
\mathcal{A}^{ABCD}_{4,\mathrm{gluon-loop}}(s,t)=&-\frac{1}{6}\left(\frac{\mathtt{g}_{\text{8d}}}{4\pi}\right)^4\,(u+s/w)^2\\
&\times \left[\mathtt{d}_{s t}^{ABCD} f_{\mathrm{box}}(s, t)+\mathtt{d}_{s u}^{ABCD} f_{\mathrm{box}}(s, u)+\mathtt{d}_{t u}^{ABCD} f_{\mathrm{box}}(t, u)\right]\,,
}
respectively, where $f_{\text{box}}(s,t)$ is the function introduced in \eqref{f_box}. Using \eqref{kappagYM}, we see that these results do indeed reproduce \eqref{Atree} and \eqref{Aloop}, respectively. In the expressions above we have used $w$ to denote the ratio of 8d gluon polarizations
\es{}{
w=\frac{(\epsilon_1\cdot \epsilon_2)(\epsilon_3\cdot \epsilon_4)}{(\epsilon_1\cdot \epsilon_3)(\epsilon_2\cdot \epsilon_4)}\,,
}
which is the flat space analogue of the R-symmetry cross ratio introduced in \eqref{crossratios}, since R-symmetry polarizations $y$ reduce to gluon polarizations $\epsilon$ in the flat space limit.

The other amplitude that we need is that corresponding to the exchange of a 10d graviton between four 8d gluons. As first discussed in \cite{Chester:2023qwo}, this can be determined by unitarity cuts and, since gluons only propagate in eight dimensions while gravitons can explore the bulk, it involves an integral over the momenta along the two directions transverse to the sevenbranes. The result can be written as
\es{A4grav}{
\mathcal{A}^{ABCD}_{4,\mathrm{grav-exch}}(s,t)=\frac{\Delta}{2}\kappa^2_{\mathrm{10d}}(u+s/w)^2\,\left[\delta^{AB}\delta^{CD}\int_{\R^2}\frac{d^2p_{\perp}}{(2\pi)^2}\frac{1}{-s+p^2_{\perp}}+\mathrm{perm.}\right]\,,
}
where $\kappa_{\mathrm{10d}}$ is the gravitational coupling constant and the result is proportional to the coefficient $\Delta$ introduced in Section \ref{sec:theories}. This is in agreement with the result of \cite{Chester:2023qwo}, where the authors study M-theory on the orbifold AdS$_4\times S^7/\Z_k$ and the graviton exchange amplitude turns out to be proportional to $k$. Here, for the orbifold cases (all theories in \eqref{sing} except $A_1$ and $A_2$), the theory is type IIB on AdS$_5\times S^5/\Z_{\Delta}$, so in a similar way we find that the graviton exchange amplitude is proportional to $\Delta$. We also note that for $\Delta=2$, which is the case for the $D_4$ theory, this reproduces the graviton exchange amplitude given in \cite{Behan:2023fqq}. Using the relation  \eqref{kappagYM} between $\kappa_{\mathrm{10d}}$ and the string theory parameters, it is straightforward to see that \eqref{A4grav} reproduces the result \eqref{AR} quoted in Section \ref{sec:mellin}.

\section{Modular functions}\label{app:modular}

In this appendix we collect some useful definitions and identities that we used in the main text. We begin defining Jacobi theta functions via
\es{}{
\theta_2(\tau)&=\sum_{n=-\infty}^{\infty} q^{(n+1 / 2)^2}=2 q^{1 / 4} \prod_{n=1}^{\infty}\left(1-q^{2 n}\right)\left(1+q^{2 n}\right)^2\,, \\
\theta_3(\tau)&=\sum_{n=-\infty}^{\infty} q^{n^2}=\prod_{n=1}^{\infty}\left(1-q^{2 n}\right)\left(1+q^{2 n-1}\right)^2\,, \\
 \theta_4(\tau)&=\sum_{n=-\infty}^{\infty}(-1)^n q^{n^2}=\prod_{n=1}^{\infty}\left(1-q^{2 n}\right)\left(1-q^{2 n-1}\right)^2\,,
}
where
\es{}{
q=e^{i\pi\tau}\,,
}
which are related to the Dedekind eta function
\es{}{
\eta(\tau)=\sum_{n=-\infty}^{\infty}(-1)^n q^{3(n-1 / 6)^2}=q^{\frac{1}{12}} \prod_{n=1}^{\infty}\left(1-q^{2 n}\right)\,,
}
via
\es{}{
 \theta_2(\tau) \theta_3(\tau) \theta_4(\tau)=2\eta(\tau)^3\,.
}
An important identity satisfied by Jacobi theta functions which we have used in the derivation of $\tau^{(2)}$ and $\omega_1^{(0)}$ in Section \ref{sec:localization} is the Jacobi identity
\es{jacobi_identity}{
\theta_2(\tau)^4+\theta_4(\tau)^4=\theta_3(\tau)^4\,.
}

In the discussion of SW theory in Section \ref{sec:localization} we have also introduced the Eisenstein series 
\es{}{
E_k(\tau)=1+\frac{2}{\zeta(1-k)} \sum_{n=1}^{\infty} n^{k-1} \frac{q^{2 n}}{1-q^{2 n}}\,,
}
where $\zeta$ is the Riemann zeta function. In particular we have used $E_4$ and $E_6$, which together with $E_2$ form a basis of quasi-modular forms and they can be expressed in terms of Jacobi theta functions via
\es{E_from_theta}{
E_2(\tau)&=\frac{12}{i \pi}\partial_{\tau}\log\eta(\tau)\,,\\
E_4(\tau)&=\frac{1}{2}(\theta_2(\tau)^8+\theta_3(\tau)^8+\theta_4(\tau)^8)\,,\\
E_6(\tau)&=\sqrt{E_4(\tau)^3-1728\eta(\tau)^{24}}\,.
}
In Section \ref{sec:localization} we have also used the following identities for the derivatives of Eisenstein series
\es{eiseinstein_derivative}{
E_2^{\prime}(\tau)&=\frac{\pi i}{6}\left[E_2(\tau)^2-E_4(\tau)\right]\,,\\
E_4^{\prime}(\tau)&=\frac{2 \pi i}{3}\left[E_2(\tau) E_4(\tau)-E_6(\tau)\right]\,, \\
E_6^{\prime}(\tau)&=\pi i\left[E_2(\tau) E_6(\tau)-E_4(\tau)^2\right]\,,
}
as well as the following values of $E_4$ and $E_6$ at special points
\es{einsenstein_fixedtau}{
E_4(i)&=\frac{3 \pi^2}{4 \Gamma(3 / 4)^8}\,, \quad E_6(i)=0\,, \\
E_4\left(e^{\frac{\pi i}{3}}\right)&=0\,, \hspace{1.2cm} E_6\left(e^{\frac{\pi i}{3}}\right)=\frac{27 \Gamma(1 / 3)^{18}}{512 \pi^{12}}\,.
}

\providecommand{\href}[2]{#2}\begingroup\raggedright\endgroup

\end{document}